\documentclass[aps,amsmath,twocolumn,amssymb,floatfix,showpacs,superscriptaddress,nofootinbib,longbibliography]{revtex4-1}
\usepackage{amsmath}
\usepackage{mathtools}
\usepackage{braket}
\usepackage{graphicx}
\usepackage{amssymb}
\usepackage[dvipsnames]{xcolor}
\usepackage{multirow}
\usepackage{xfrac}
\usepackage{textcomp}
\usepackage{float}
\usepackage{enumerate}
\usepackage{subfigure}
\usepackage{stackengine}
\usepackage{booktabs}
\usepackage{physics}
\usepackage[colorlinks=true,linktoc=page,linkcolor=PineGreen,citecolor=magenta,urlcolor=violet]{hyperref}
\usepackage{verbatim}

\newcommand{\vect}[1]{\boldsymbol{\mathrm{#1}}}
\mathchardef\mhyphen="2D 

\newcommand{\etal}{{\it et al.~}}
\newcommand{\ie}{{\it i.e.,\,\,}}
\newcommand{\eg}{{\it e.g.,~}}

\newcommand\bea{\begin{eqnarray}}
\newcommand\eea{\end{eqnarray}}
\newcommand\beq{\begin{equation}}  
\newcommand\eeq{\end{equation}}

\newcommand{\non}{\nonumber}  
\usepackage[normalem]{ulem}
\definecolor{lime}{HTML}{A6CE39}
\usepackage{sidecap,tikz}
\DeclareRobustCommand{\orcidicon}{\hspace{-1.0mm}
	\begin{tikzpicture}
	\draw[lime, fill=lime] (0.0,0.0) 
	circle [radius=0.15] 
	node[white] {{\fontfamily{qag}\selectfont \tiny \,ID}};
	\draw[white, fill=white] (-0.0525,0.095) 
	circle [radius=0.007];
	\end{tikzpicture}
	\hspace{-3.0mm}
}
\foreach \x in {A, ..., Z}{\expandafter\xdef\csname orcid\x\endcsname{\noexpand\href{https://orcid.org/\csname orcidauthor\x\endcsname}
		{\noexpand\orcidicon}}
}

\begin{document}
	

\title{Topological Superconductivity by Engineering Noncollinear Magnetism in Magnet/ Superconductor Heterostructures: A Realistic Prescription for 2D Kitaev Model}  

\author{Pritam Chatterjee\orcidA{}}
\altaffiliation{These authors contributed equally to this work}
\affiliation{Institute of Physics, Sachivalaya Marg, Bhubaneswar-751005, India}
\affiliation{Homi Bhabha National Institute, Training School Complex, Anushakti Nagar, Mumbai 400094, India}
\author{Sayan Banik\orcidH{}}
\altaffiliation{These authors contributed equally to this work}
\affiliation{School of Physical Sciences, National Institute of Science Education and Research, An OCC of Homi Bhabha National Institute, Jatni 752050, India}
\author{Sandip Bera\orcidD{}}
\affiliation{School of Physical Sciences, National Institute of Science Education and Research, An OCC of Homi Bhabha National Institute, Jatni 752050, India}
\author{Arnob Kumar Ghosh\orcidC{}}
\affiliation{Institute of Physics, Sachivalaya Marg, Bhubaneswar-751005, India}
\affiliation{Homi Bhabha National Institute, Training School Complex, Anushakti Nagar, Mumbai 400094, India}
\author{Saurabh Pradhan\orcidE{}}
\affiliation{Lehrstuhl f\"ur Theoretische Physik II, Technische Universit\"at Dortmund Otto-Hahn-Str. 4, 44221 Dortmund, Germany}
\author{Arijit Saha\orcidF{}}
\email{arijit@iopb.res.in}
\affiliation{Institute of Physics, Sachivalaya Marg, Bhubaneswar-751005, India}
\affiliation{Homi Bhabha National Institute, Training School Complex, Anushakti Nagar, Mumbai 400094, India}
\author{Ashis K. Nandy\orcidG{}}
\email{aknandy@niser.ac.in}
\affiliation{School of Physical Sciences, National Institute of Science Education and Research, An OCC of Homi Bhabha National Institute, Jatni 752050, India}

\begin{abstract} \noindent
We report on a realistic and rather general scheme where noncollinear magnetic textures proximitized with the most common $s$-wave superconductor can appear as the alternative to $p$-wave superconductor{--}the prime proposal to realize two-dimensional (2D) Kitaev model for topological superconductors (TSCs) hosting Majorana flat edge mode (MFEM). A general minimal Hamiltonian 
suitable for magnet/superconductor heterostructures reveals robust MFEM within the gap of Shiba bands due to the emergence of an effective ``$p_x+p_y$"-type $p$-wave pairing, spatially localized at the edges of a 2D magnetic domain of spin-spiral. We finally verify this concept by considering Mn (Cr) monolayer grown on a $s$-wave superconducting substrate, Nb(110) under strain (Nb(001)). In both 2D cases, the antiferromagnetic spin-spiral solutions exhibit robust MFEM at certain domain edges that is beyond the scope of the trivial extension of 1D spin-chain 
model in 2D. This approach, particularly when the MFEM appears in the TSC phase for such heterostructure materials, offers a perspective to extend the realm of the TSC in 2D.
\end{abstract}
\maketitle
\textcolor{blue}{\emph{Introduction.---}} 
A strong quest for topological superconductors~(TSCs) hosting Majorana zero-modes (MZMs)~\cite{Kitaev2001,Ivanov2001,SDSarma2008,Kitaev2009,qi2011topological,Alicea_2012,Leijnse_2012,beenakker2013search,ramonaquado2017} has been accumulating an immense interest based on magnetic adatoms fabricated on top of an $s$-wave superconductor (SC) substrate~\cite{Felix2013,AliYazdani2013,DanielLoss2013,PascalSimon2013,MFranz2013,Eugene2013,Felix2014, TeemuOjanen2014,MFranz2014,Rajiv2015,Sarma2015,Hoffman2016,Jens2016,Tewari2016,
PascalSimon2017,Theiler2019,Cristian2019,Mashkoori2019,Menard2019,Pradhan2020,Teixeira2020,
Alexander2020,Perrin2021}. These magnetic atoms in the presence of superconductivity lead to the formation of Yu-Shiba-Rusinov (YSR)/Shiba bands~\cite{Shiba1968,Felix2013,AliYazdani2013} inside the superconducting gap. The mini gap created within such bands plays a pivotal role in exhibiting topological MZMs~\cite{Felix2013,AliYazdani2013,Kaladzhyan2016,Yong2022,Sebastian2022,
Ghazaryan2022,Schmid2022,Pritam2023} through phase transitions, akin to the one-dimensional Kitaev model (1D-KM)~\cite{Kitaev2001,Kitaev2009}. Generally, the corresponding features like the Shiba states and/or the MZMs are experimentally detected~\cite{Eigler1997,Yazdani1999,Yazdani2015,Wiesendanger2021,Beck2021,Wang2021,Schneider2022}
in an 1D spin-chain mimicking a trail of magnetic impurities when grown on an $s$-wave SC. It is essential to highlight that a transition from a 1D finite ferromagnetic (FM) spin chain model with Rashba SOC to its 2D counterpart (FM finite domain) reveals fresh and unique phenomena: the TSC phase in 1D is manifested through the emergence of MZMs \cite{LiPRB2014,BrydonPRB2015} while 
a generalized ``$p_x+ip_y$"-type pairing governing 2D TSC phase brings higher Chern numbers and dispersive chiral Majorana edge modes~\cite{Ojanen12015,Ojanen22016}. Preserving all essential terms in those models, this dimensional extension in the problem introduces unique and distinct physics beyond the scope of its 1D counterpart. Therefore,
moving to 2D-KM, the prime proposal turns out to be the $p$-wave SCs and thus, there has been a growing consensus on realizing $p$-wave SCs in materials despite its rarity so far. A distinct signature of such TSCs is non-dispersive Majorana flat edge modes~(MFEMs) localized at edges of a 2D domain, can be probed experimentally using scanning tunnelling microscopy (STM) and  angle resolved photo emission spectroscopy (ARPES) .
 
\begin{figure}[]
	\centering
	\subfigure{\includegraphics[width=0.49\textwidth]{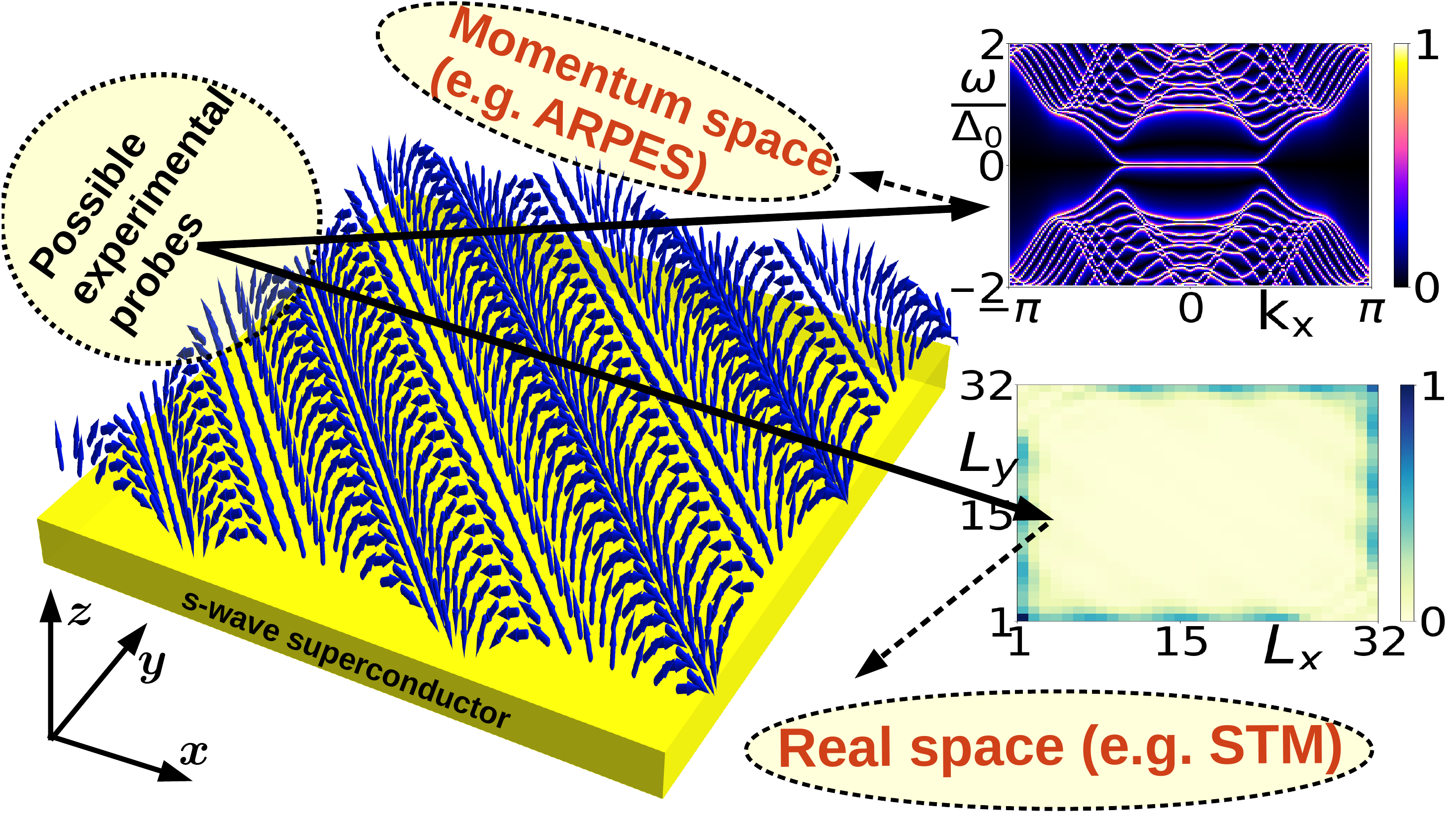}}
	\vspace{-0.4 cm}
	\caption{The schematic setup of our model, a 2D square lattice with SS state placed on the surface of an $s$-wave SC. Two possible experimental schemes to probe MFEM: the momentum space probe (\eg ARPES), measures the spectral function as a signature of the MFEM and the real space probe (\eg STM), measures the spatial distribution of the MFEM at edges, see the corresponding right images.}
	\label{TOYFIG}
	\vspace{-0.5 cm}
\end{figure}

Recently, theoretical proposals for the 2D-KM with topological gapless phase hosting MFEMs has been put forward by employing $(p_x$+$p_y)$-SCs~\cite{Wang2017,Zhang2019}. A few alternative schemes using inhomogeneous magnetic fields, various magnetic orders $etc$ were also explored to generate different $p$-wave pairing~\cite{Bena2015,Nagaosa2013,Chen2015}. Although a formal connection involving both model and real materials, manifesting similar behavior has never been proposed in this context. Hence, we can address the following intriguing questions that have not been answered so far to the best of our knowledge: (a) Can we architect and identify magnetic heterostructures where the spin-spiral (SS) solution in the presence of $s$-wave SC exhibits features of 2D-KM? (b) Is it possible to derive an effective continuum model consisting of an induced effective spin-orbit coupling (SOC) and Zeeman field to describe such system? (c) Finally and the most importantly, can we identify prototype systems where the SS ground state exhibits gapless TSC phase hosting MFEM within a lattice model? By stabilizing the SS state in 2D films comprising of 3$d$ transition metal (TM) monolayer and $s$-wave SC substrate may offer the most promising platform for stabilizing the TSC phase in experiments.

In this letter, we first deal with the SS textures in 2D, mimicking spatially varying magnetic impurities proximitized with a $s$-wave SC. An effective $(p_x + p_y)$-SC pairing is identified for the SS propagating along [110] direction in a square domain, manifesting gapless TSC phase. Note, the signature of the TSC phase \ie the nondispersive MFEMs cannot be obtained by straightforward generalization of 1D system~\cite{AliYazdani2013}. 
The interplay between the SS sate and $s$-wave SC in our minimal model leads to the emergence of a distinctive ``$p_x+p_y$" type SC pairing, supporting the existence of a gapless topological superconducting phase in 2D. At the end, the designed TM/SC heterostructures must reveal SS ground state. We design potential material candidates: one monolayer of Mn and Cr on Nb(110) and Nb(001) SC substrates, respectively. Experimentally observed in-gap YSR band in Mn/Nb(110)~\cite{Conte2022} is well reproduced within our minimal model, owing to the proximity induced SC in the antiferromagnetic (AFM) state. By looking other aspects of 2D noncollinear magnets \cite{Nandy2016,NandyNanoLett20,Bode_Nat07}, we apply uniform biaxial strain for engineering AFM-SS state as the ground state in Mn/Nb(110). The AFM-SS within a lattice model reveals TSC phase hosting MFEM. The TSC phase is further observed in another example, unstrained Cr/Nb(001). Hence, such real materials platform adds significant merit to the problem we are dealing with.


\textcolor{blue}{\emph{Formulation of 2D Kitaev continuum model}.---}
Within a continuum model, we first propose a general route to design 2D gapless TSC phase via engineering SS textures, when proximitized with an $s$-wave SC. The 2D model Hamiltonian for locally varying magnetic impurities reads in the Nambu spinor form,
$\Psi(\vect{r})$=$\begin{pmatrix} c_{\vect{r},\uparrow},\! & \! c_{\vect{r},\downarrow},\! &\! c^{\dagger}_{\vect{r},\downarrow},\! &\!\! -c^{\dagger}_{\vect{r},\uparrow} \end{pmatrix}^{\rm T}\!$  as $\mathcal{H}_{2{\rm D}}$=$\int \! d \vect{r}\Psi^{\dagger}(\vect{r})~H~ \Psi(\vect{r})$, where $c_{\vect{r},\uparrow (\downarrow)}$ represents the quasiparticle annihilation operator for the up (down) spin at $\vect{r}$=$(x,y)$. The first quantized form of this Hamiltonian reads,
\vspace{-0.25 cm}
\begin{align}
H=& -\frac{1}{2}\vect{\nabla}^2 \tau_{z}-J \vect{S}(\vect{r}) \cdot \boldsymbol{\sigma} +\Delta_{0}\tau_{x}-\mu\tau_{z} \ . 
\label{1qeffHamil}
\end{align}
For simplicity, we consider $\hbar$=$1$ and $m$=$1$. The Pauli matrices $\sigma$ and $\tau$ acts on the spin and particle-hole subspace, respectively. $J$, $\Delta_{0}$ and  $\mu$ denote the local exchange-interaction strength between the magnetic impurity spin and electrons in the SC, the $s$-wave order parameter and the chemical potential, respectively. We assume the impurity spins to be classical and confined in the $xy$-plane with magnitude $\lvert \vect{S} \rvert$=$1$. Therefore, spin vector $\vect{S}(\vect{r})$ can be locally described as $\vect{S}(\vect{r})$=$\lvert \vect{S} \rvert\begin{pmatrix} \cos[\phi(\vect{r})],\!&\! \sin[\phi(\vect{r})],\! &\! 0 \end{pmatrix}$ with $\phi(\vect{r})$, the angle between two adjacent spins. By a unitary transformation $U$=$e^{-\frac{i}{2}\phi(\vect{r})\sigma_z}$, an effective low-energy Hamiltonian $\tilde{H}$=$U^\dagger H U$ in 2D becomes,
\vspace{-0.2 cm}
\begin{align}
	\tilde{H}\!=\! &- \frac{1}{2} \! \! \sum_{r_i=x,y} \!\! \left[ \! \nabla_{r_i}^2 \! - \!\left(\!\pdv{\phi}{r_i}\!\right)^{\!2}\!\!- \! \left(\! i\pdv{\phi}{r_i}\nabla_{r_i} \!+ i\nabla_{r_i} \pdv{\phi}{r_i}\right) \! \sigma_z \! \right] \! \tau_z   \nonumber \\
  & ~~~~~~~~~~~~~~~~~~~~~~~~~~~~~~~~~ - \! J\sigma_x \! + \! \Delta_0 \tau_x \! - \!  \mu\tau_z \ .
	\label{eq:TransformedHelicalChain}
\end{align}
Note, the similar mapping was reported for 1D spin chain model in case of Majorana bound state solution~\cite{DanielLoss2022}. 

Henceforth, we assume that the SS is propagating along the diagonal of a square domain as depicted  in Fig.~\ref{TOYFIG}, effectively along [110] direction. The angle, $\phi(\vect{r})$=$\vect{g}\!\cdot \!\vect{r}$=$\left( g_x x+g_y y \right)$, defines the angle between two adjacent spins along the SS propagation direction where $g_x $ and $g_y$ values control the SS period and propagation direction. Generally, for $\vert g_x\rvert \!\! \ne \!\! \vert g_y \rvert$, one finds an asymmetric spin textures where the SS propagates neither [110] nor [1$\bar{1}$0] directions and the Hamiltonian $\tilde{H}$ in Eq.~(\ref{eq:TransformedHelicalChain}) can be rewritten in the momentum space as

\begin{eqnarray}
\tilde{H} (\vect{k})\!=\!&\xi_{\vect{k},\vect{g}}\tau_{z}\! +\! \frac{1}{2} \vect{g} \! \cdot \! \vect{k} \ \sigma_z\tau_{z} \! +\! J\sigma_x \! + \! \Delta_{0}\tau_{x} \ ,  
\label{modifiedhamiltonian}
\end{eqnarray}

where, $\xi_{\vect{k},\vect{g}}$=$\frac{1}{2}\left( \vect{k}^2\!+\!\vect{g}^2 \right)\!-\!\mu$. 
The second term represents an effective SOC, resulting from the spin texture in our model.
Although, the nature of such SOC is quite nontrivial as it originates from the spin texture, it can interestingly show a gapless TSC phase in the presence of Zeeman like field of strength $J$ along the $x$-direction. 
We obtain the spectrum for the Hamiltonian $\tilde{H} (\vect{k})$ Eq.~(\ref{modifiedhamiltonian}) as, 
$E_{r,s}(\vect{k},\vect{g})$=$r \sqrt{J^{2}\!+\!\Delta_{0}^{2}\!+\!\xi_{\vect{k},\vect{g}}^{2}\!+\!\frac{1}{4} \vect{k} \cdot \vect{g} + sF(\vect{k},\vect{g})}$; where, $r,s$=$\pm$ and $F(\vect{k},\vect{g})$=$\sqrt{[(\vect{k} \cdot \vect{g})^{2}\!+\!4J^{2}]^{2}\xi_{k,g}^{2}\!+\!4J^{2}\Delta_{0}^{2}}$.
Following the gap closing condition corresponding to the two lowest energy bands, the critical value of $J$ becomes 
$J_{c}(\vect{g})$=$\sqrt{\Delta_0^2 \! + \! \left(\mu \! - \! \vect{g}^2/2\right)^2}$.
\label{critfield}
The SS period now can be manifested like $T$=$\pi/\vert \vect{g} \rvert$=$\pi/\! \sqrt{g_x^2\! + \! g_y^2}$~\cite{DanielLoss2022}. Hence, in case of $\vert g_x\rvert$=$\vert g_y\rvert$=$g$, the period turns out to be $T$=$\pi/\sqrt{2}g$. Naively,  the SS solution is governed by the RKKY-type (Ruderman-Kittel-Kasuya-Yosida) exchange frustration and if the period of SS is set by the Fermi momentum $\vect{k}_F$, then $T$=$\pi/\lvert \vect{g}\rvert$=$\pi/\lvert \vect{k}_F \rvert$~\cite{DanielLoss2022,DanielLoss2013,PascalSimon2013,MFranz2013}. In such case, the topological transition occurs at $J_{c}$=$\Delta_0$ so that $\mu$=$\vert \vect{k}_F \rvert^2/2$. However, in real TM/SC systems, the SS solution is the outcome of a complex interplay of material dependent parameters: the exchange coupling constants, $\mathcal{J}_{ij}$'s; the Dzyaloshinskii-Moriya Interactions (DMIs), $\textbf{D}_{ij}$'s;  and the uniaxial magneto crystalline anisotropy, $\mathcal{K}$.  

\begin{figure}[]
	\begin{center}
		\includegraphics[width=0.45\textwidth]{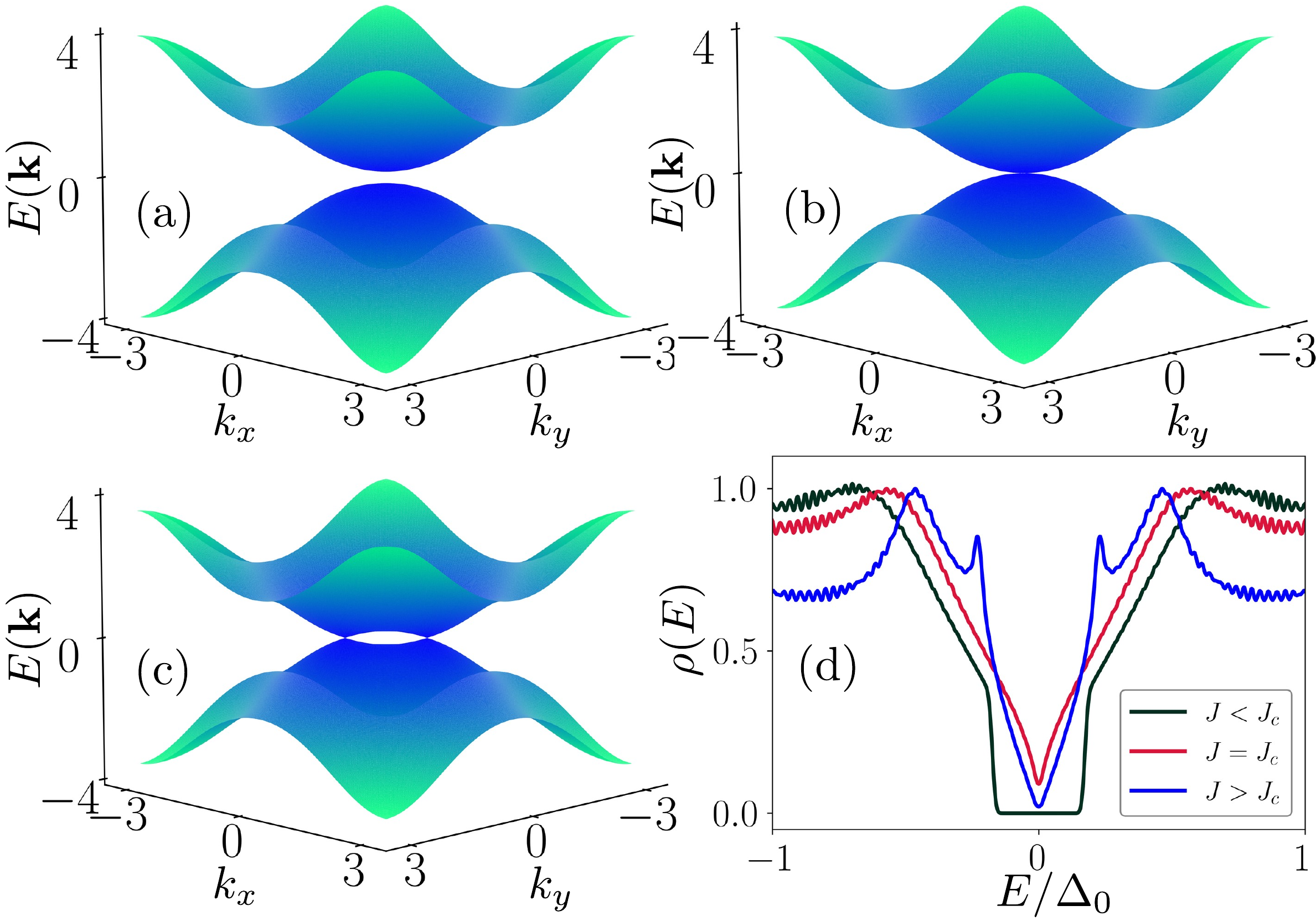}
	\vspace{-0.5 cm}
	\end{center}
	\caption{The bulk band structures of the Hamiltonian $\tilde{H}_{\rm L}(\vect{k})$ in the $k_x$-$k_y$ plane is depicted for (a) the trivially gapped SC phase, $J<J_c$ (=1.6$\Delta_0$), (b) the gap closing transition point, $J=J_c$ (=1.77$\Delta_0$), and (c) the gapless TSC phase hosting MFEM, $J>J_c$ (=2.0$\Delta_0$). (d) TDOS, $\rho(E)$, is shown as a function of $E/\Delta_0$ for the above mentioned $J$ values in the parenthesis. All remaining parameters take values: $g_x$=$g_y$=$\pi/2$, 
$\mu$=$\Delta_0=t$.} 
	\label{bands}
	\vspace{-0.5 cm}
\end{figure}

The band structure of this system has been analyzed using the lattice version of the Hamiltonian $\tilde{H}(\vect{k})$ defined as $\tilde{H}_{\rm L}(\vect{k})$, see Eq.~(S1) in the supplementary materials (SM),  Section~S1 ~\cite{supp}. We depict the bulk band structure of $\tilde{H}_{\rm L}(\vect{k})$ and the corresponding total density of states~(TDOS) $\rho(E)$ in Fig.~\ref{bands}. 
The topological phase transition occurs between a normal SC phase with a trivial gap, Fig.~\ref{bands}(a) for $J \!\! <\!\! J_c$ to the gapless TSC phase,  Fig.~\ref{bands}(c) for $J \!\!>\!\! J_c$ via a gap closing phase, Fig.~\ref{bands}(b) at $J$=$J_{c}$ =1.77$\Delta_0$. The topological characterization via appropriate topological invariant ($\nu$) is provided in the SM, Section~S1~\cite{supp}. The invariant $\nu$ changes from $0$ to $1$ and hence, the system undergoes a transition from a trivial gaped state ($J \!\!<\!\! J_c$) to a non-trivial ($J \!\!>\!\! J_c$) TSC phase. This gapless phase displays graphene like semimetalic behavior~\cite{Castro2009,Wakabayashi2010} where $\rho(E)$ corresponding to the gapless TSC phase ($J \!\!>\!\! J_c$) varies almost linearly with $E$, see Fig.~\ref{bands}(d). The band structure illustrated in Fig.~\ref{bands}(c) resembles that of the 2D-KM in the TSC phase, reported recently in Refs.~\cite{Zhang2019,Wang2017}. There, the idea of 2D-KM hosting MFEM has been analytically formulated considering  ($p_x+p_y$)-SCs. Indeed, we derive an effective ``$p_x+p_y$" SC pairing [based on Eq.~(\ref{modifiedhamiltonian})] as a result of a domain of SS states propagating along the [110] direction when it is proximitized with an $s$-wave SC, see details in Section S6 in the SM~\cite{supp}. Moreover, the inclusion of Rashba SOC in our model and the extension to multi-orbitals~\cite{HughesTaylor2008,AndersMathias2016} (see Section S7 in the SM~\cite{supp}) do not significantly affect the presence and characteristics of MFEMs. These results seemingly ensure that the crucial prerequisite for the TSC phase is the noncollinear SS state stabilized in TM/SC systems.

\begin{figure}[]
	\begin{center} 
		\includegraphics[width=0.45\textwidth]{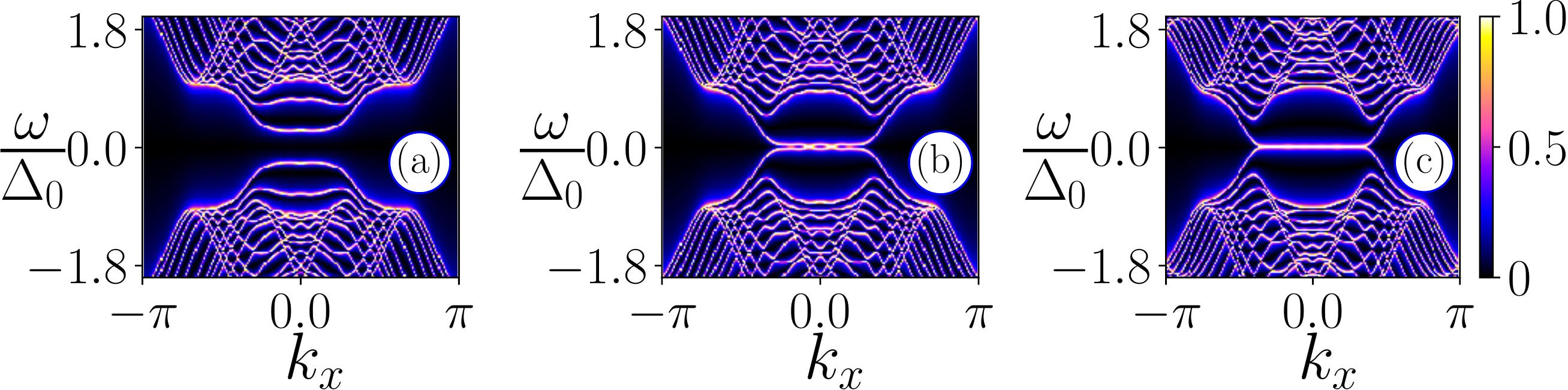}
		\vspace{-0.20 cm}
	\end{center}
	\vspace{-0.5cm}
	\caption{Panels (a)-(c), the density plots of $\mathcal{A}(k_x,\omega)$ in the $k_x$-$\omega$ plane. A signature of the MFEM is  seen in panel (c) for $J\!>\!J_c$ (= 2.0$\Delta_0$) where the bulk is a gapless TSC. We choose the same set of respective parameters as mentioned in Fig.~\ref{bands}.}
	\label{spectral function}
	\vspace{-0.5 cm}
\end{figure}

In a slab geometry, we then calculate the spectral function, $\mathcal{A}(k_x,\omega)$~\cite{spec_func}.  Fig.~\ref{spectral function} shows the behavior of $\mathcal{A}(k_x,\omega)$ as a function of energy, $\omega/\Delta_0$. Indeed, the MFEM signature is found clearly in the gapless TSC phase in Fig.~\ref{spectral function}(c). Fig.~\ref{spectral function}(a) shows a trivial gap, \ie without any signature of MFEM and in Fig.~\ref{spectral function}(b) for $J$=$J_{c}$, the edge modes in $\mathcal{A}(k_x,\omega)$ plot are still infinitesimally gaped. Experimentally, one can probe these signatures of TSC phase using ARPES measurements but such a small gap close to the transition point will be impossible to resolve. 

\begin{figure}[]
	\begin{center}
		\includegraphics[width=0.5\textwidth]{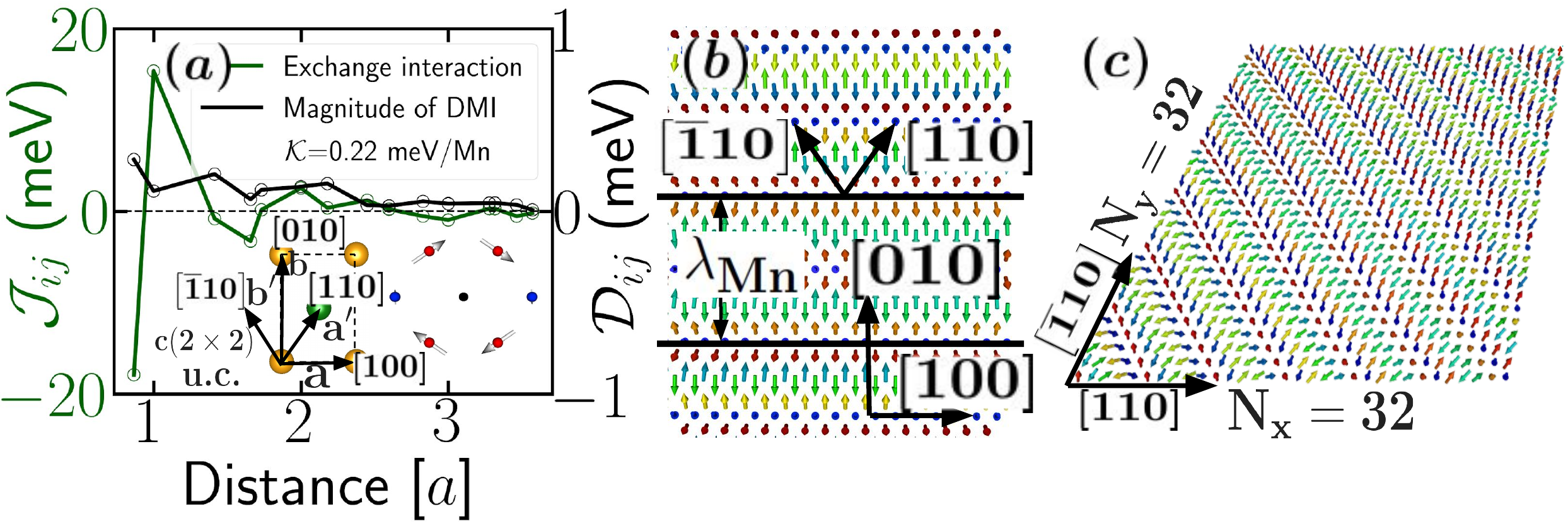}
	\end{center}
	\vspace{-0.65 cm}
	\caption{For Mn/Nb(110) with strain $\sim \!\! -2.7$\%, (a) $\mathcal{J}_{ij}$'s and $\mathcal{D}_{ij}$'s are plotted as a function of distance measured in units of lattice constant $a$. Calculated $\mathcal{K}$ is out-of-plane. Lower left inset shows the $c(2\times2)$-AFM surface unit cell (yellow and green balls represent up and down spins, respectively), possessing $\mathcal{C}_{2v}$ symmetry. The lower right inset describes vector-orientations of DMIs. (b) The AFM-SS ground state propagates along the [010] direction. (c) A $32\times 32$ AFM-SS domain where edges are considered along [110] and [$\bar{1}$10] directions.}
\label{theta_phi}
\vspace{-0.4 cm}
\end{figure}

\textcolor{blue}{\emph{2D Kitaev lattice model for TM/SC heterostructure.---}}
Focusing on realistic materials framework, we rationalize the above described phenomena using their magnetic ground states.  We, therefore, design two prototype TM/SC heterostructures based on 3$d$-TM monolayer grown on $s$-wave SC substrates: Mn/Nb(110) and Cr/Nb(001). 
The Mn/Nb(110) example with its relaxed film geometry constructed with the optimized lattice constant of bulk Nb~\cite{opt_lat} indeed show a $c(2 \times 2)$ AFM order as the ground state (for detail results, see SM, Section~S3(A)~\cite{supp}), recently reported in experiment also~\cite{Conte2022}. Surprisingly, we find a transition to an AFM-SS state via a uniform biaxial compressive strain within the range, $\sim-1$ to $-4$ \%. Considering $a$=3.234 \AA~and $b$=$\sqrt{a}$=4.574~\AA~(strain \! $\sim \! -2.7$~\%), Fig.~\ref{theta_phi}(a) illustrates $\mathcal{J}_{ij}$'s and the absolute values of DMI ($\mathcal{D}_{ij}$) as a function of distance between Mn atoms and here, $\mathcal{K}$ is positive \ie out-of-plane. The vector orientations of DMIs in the inset connecting neighboring atoms match the symmetry rules for a system with $\mathcal{C}_{2v}$ symmetry~\cite{DMI2}. Calculational details and more results with varying planner strains on Mn/Nb(110) film are provided in the SM, Section~S3(B)~\cite{supp}. The AFM-SS solution occurs as the stable state even without DMIs, resulting from the strong frustration in $\mathcal{J}_{ij}$'s connecting Mn moments ($\mathcal{M}_{\textrm{Mn}}$= 3.53~$\mu_\textrm{B}$). The significantly weak DMI strengths ($< \! 0.5~meV$) are attributed to the weak SOC in light atoms and here, it determines a right-handed cycloidal AFM-SS as the ground state propagating along the [010] direction, see Fig.~\ref{theta_phi}(b). 

Here, we elucidate a minimal electronic model Hamiltonian in real space for a 2D lattice,
\vspace{-0.25 cm}
\begin{eqnarray}
H&=& \! -\!\!\sum_{\langle i,j \rangle,\alpha}  t_{ij} c_{i,\alpha}^{\dagger}  c_{j,\alpha} \! -\! \mu \sum_{i,\alpha} c_{i,\alpha}^{\dagger} c_{j,\alpha}   \nonumber\\
&&+J \! \! \sum_{i,\alpha,\beta} \! \! c_{i,\alpha}^{\dagger} \! \left(\vect{\hat{s}}_i \! \cdot \! \vect{\sigma}\right)_{\alpha, \beta} c_{i,\beta}\!+\! \Delta_{0}\! \sum_{i}
(c_{i,\uparrow}^{\dagger}c_{i,\downarrow}^{\dagger}\!+\! {\rm H.c.})\ , \quad ~
\label{NuModel}\ 
\end{eqnarray}
where, $i,j$ indices run over all the lattice sites, $\alpha,\beta$ denote the spin, $\langle \rangle$ represents nearest-neighbor hopping only, $\mu$, $J$, $\Delta_{0}$ represent the chemical potential, exchange coupling strength, $s$-wave SC gap, respectively and $c^\dagger$($c$) corresponds to the electron creation (annihilation) operator for the SC. For simplicity, we assume the hopping amplitude $t_{ij}$=$t_{ji}$=$-t$ with $t$=1 for the overall energy scale of our system. This minimal Hamiltonian is essentially constructed to ensure the importance of magnetic textures (alternative to the intrinsic SOC) of TM/SCs in the context of stabilizing TSC phase. All spin textures are actually entered in the third term in Eq.~(\ref{NuModel}), describing a local interaction between the electron's spin ($\sigma$) and the moments of Mn or Cr. The unit vector $\vect{\hat{s}}_i$ denotes $(\sin\theta_{i}\cos\phi_{i},\sin\theta_{i}\sin\phi_{i},\cos\theta_{i})$, mimicking locally varying magnetic impurities. One can extract these $\theta$ and $\phi$ from the spin textures generated by the MC simulations for TM/SC systems.

In case of Mn/Nb(110), the $c(2\times2)$ AFM phase has been assessed first via numerically solving Eq.~(\ref{NuModel}). Our results  indeed describe the experimental findings where the $s$-wave SC and AFM phases are found to coexist~\cite{Conte2022}, see the SM, Section~S3(B) for more details~\cite{supp}.The in-gap YSR bands ensure the qualitative accuracy of our minimal model in Eq.~(\ref{NuModel}).

\begin{figure}[]
	\begin{center}
		\includegraphics[width=0.5\textwidth]{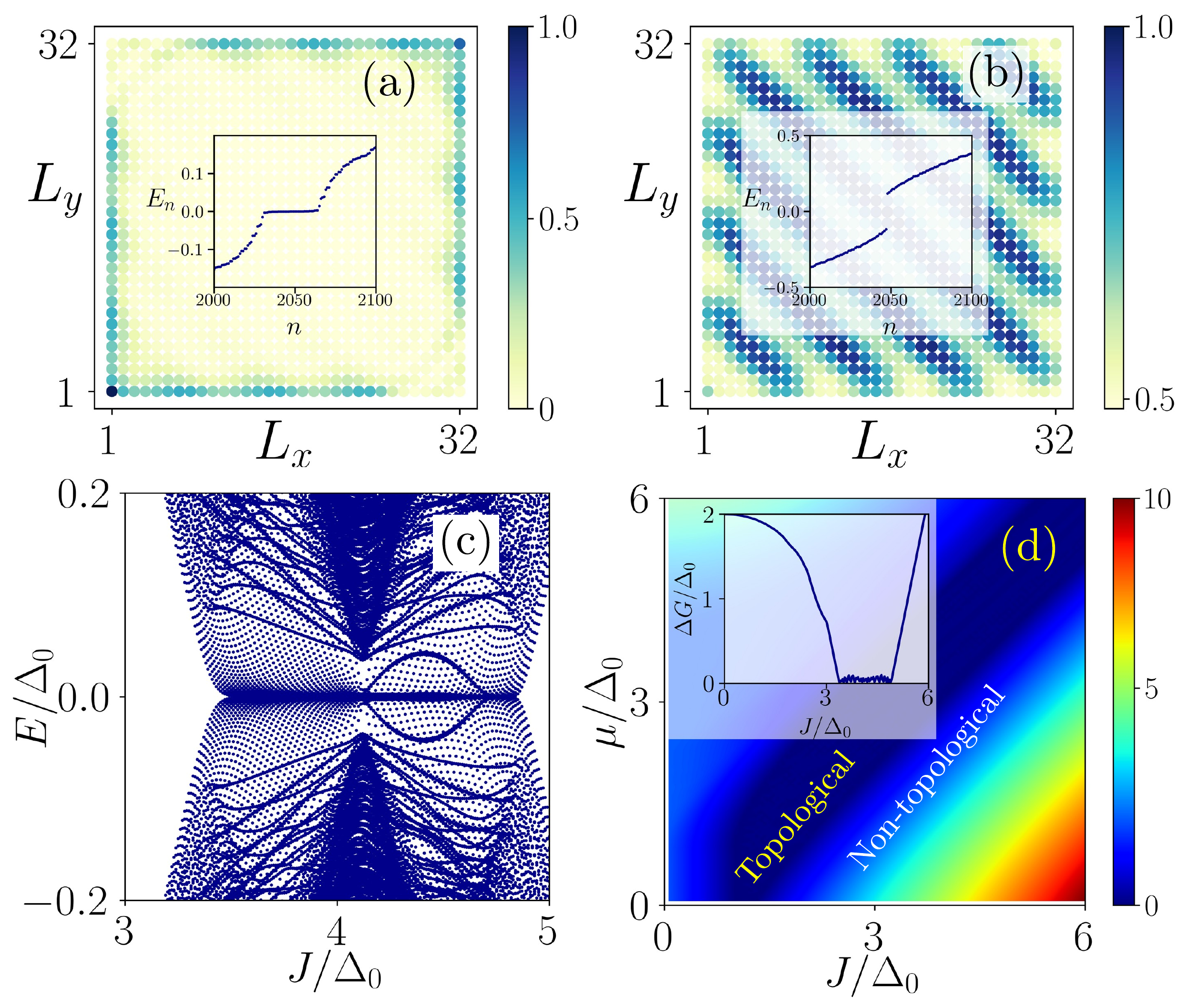}
	\end{center}
	\vspace{-0.7 cm}
	\caption{The normalized LDOS for $E$=0 eigenstate, computed within a spin lattice model of size 32$\times$32 spins ($L_x \mhyphen L_y$ square plane). Here, 2D lattice points are defined in unit of, $ \vect{a^\prime}$ and $\vect{b^\prime}$, as shown in the inset of Fig.~\ref{theta_phi}(a). For Mn/Nb(110) with AFM-SS state, (a) we identify the TSC phase for $J$= 4.5$\Delta_0$. The LDOS is predominantly localized at the domain edges denoting MFEMs. (b) In the trivial phase for $J$= 5.0$\Delta_0$, LDOS is delocalized over the entire domain. Insets in (a) and (b) show a zero-energy flat mode and a trivial gap  in the eigenvalue $E_n$ vs. state $n$ plots, respectively. (c) Energy eigenvalues $E$ of $H$ is shown as a function of $J$ using OBC. (d) The bulk-gap $\Delta G$ profile is shown in the $J \mhyphen \mu$ plane employing PBC, indicating the TSC phase (dark blue regime). The inset is showing $\Delta G$ vs. $J$ plot for a fixed $\mu$ (=4.0$\Delta_0$ and $\Delta_0$=$t$).}
	\label{MFB_LDOS}
	\vspace{-0.55 cm}
\end{figure}

A few important results are obtained from the numerical simulations by considering the AFM-SS 2D domain. We construct a domain of size 32$\times$32 (1024 spins) where the AFM-SS is propagating along the diagonal of that domain (as similar in the FM-SS, Fig.~\ref{TOYFIG}) with edges along [110] and [1$\bar{1}$0] directions, see Fig.~\ref{theta_phi}(c). Hence, these directions are parallel to the rotated vectors $ \vect{a^\prime}$ and $\vect{b^\prime}$ of the 2D lattice in the inset of Fig.~\ref{theta_phi}(a). The measured $\theta$ and $\phi$ values describe $\hat{s}$ to solve Eq.~(\ref{theta_phi}), numerically. Results are summarized in Figs.~\ref{MFB_LDOS} where, (a) and (b) depict the local density of states~(LDOS) for the zero-energy ($E$=0) states using coupling constant, $J$=4.5$\Delta_0$ and 5.0$\Delta_0$, respectively. The zero-energy states populate along the edges of the domain in Fig.~\ref{MFB_LDOS}(a) and hence, the system is in the TSC regime. The MFEM are maximally localized at the two opposite corners of the system and disperse gradually along the edges. Moreover, the signature of the MFEM is more evident from the non-dispersive states at $E_n$=0 in the eigenvalue spectrum plotted as a function of the state index $n$ in the inset of Fig.~\ref{MFB_LDOS}(a). The semimetallic behavior of the bulk YSR band at $J$ = 4.5$\Delta_0$ presented in the SM, Section S5 ~\cite{supp} for the TSC regime qualitatively matches with the continuum results presented in Fig.~\ref{bands}(d). The inset of Fig.~\ref{MFB_LDOS}(b) shows a trivial phase by opening a gap in the eigenvalue spectrum around $E_n$=0. 

Seemingly, it appears that the coupling constant $J$ (rather, $J/\Delta_{0}$) plays a major role in the phase transition between trivial SC and TSC phases. Particularly, $J$ value is often very challenging to determine for such materials. Therefore, in Fig.~\ref{MFB_LDOS}(c), we depict the eigenvalue spectrum $E/\Delta_0$ as a function of $J/\Delta_0$ by employing open boundary condition (OBC). The MFEM appears at zero-energy between $J$=3.5$\Delta_0$ and $J$=4.7$\Delta_0$, indicating the TSC regime. We thereafter identify the parameter regime where the MFEM appears via calculating the bulk-gap, $\Delta G$=$\lvert E_2$-$E_1 \rvert$, within periodic boundary condition (PBC). Here, $E_1$~($E_2$) represents the two low energy bands. 
We depict $\Delta G / \Delta_0$ in the $J / \Delta_0 \mhyphen \mu / \Delta_0$ plane in Fig.~\ref{MFB_LDOS}(d). The gapless TSC regime harboring MFEM is highlighted by the dark blue strip ($\Delta G \simeq 0$), while the regime outside ($\Delta G > 0$) represents gapped trivial superconducting phase. In the inset of Fig.~\ref{MFB_LDOS}(d), we illustrate the bulk-gap $\Delta G$ as a function of $J$ for a fixed value of $\mu$, for the transparent visibility of the TSC regime. The bulk-gap $\Delta G$ vanishes in the topological regime and MFEM appears at the boundary. 

The new example, Cr/Nb(001), shows left-handed cycloidal AFM-SS as the ground state without strain, propagating along both [100] and [010]  degenerate directions. The degenracy is owing to the symmetry rules followed by the DMI vectors in the $\mathcal{C}_{4v}$ symmetric film~\cite{DMI2}. We solve Eq.~(\ref{NuModel}) again for the square domains and all results are presented in the SM, Section S4~\cite{supp}.


\textcolor{blue}{\emph{Summary and outlook.---}}
In conclusion, by employing a continuum model, we demonstrate a route to generate a gapless TSC phase hosting MFEMs by engineering noncollinear SS state proximitized with an $s$-wave superconductor. This underlying scheme is later extended within a minimal lattice model which provides a unique way to unveil the TSC phase in prototype real TM/SC ($s$-wave) materials: Mn/Nb(110)  under strain and Cr/Nb(001). Even though, the SOC strength in Nb is expected to be very small, the AFM-SS can be stabilized from the exchange frustration, particularly in the Mn/Nb(110) sample and thereby, an effective SOC and Zeeman filed terms due to the spin textures manifest the Hamiltonian obtained in~ Eq.~(\ref{modifiedhamiltonian}).  The immobile MFEMs can be thought of edge channels to carry edge vortices in Josephson junction geometries$-$offering a possibility for non-Abelian ``braiding" operations~\cite{Alicea2011,Fatin2016}. Note that, similar idea has been demonstrated in the context of  braiding mobile vortices via the chiral dispersive edge channels in 2D~\cite{braidingPRL2019}. Importantly, the investigation of this ``braiding" protocols for MFEMs is an intriguing and important direction for future studies. Indeed, our realistic model Hamiltonian unveils excellent examples where the strain driven modulation of the noncollinear magnetic phases can offer unprecedented control over various types of TSC phases including 2D-KM, higher-order topological superconductors~\cite{SahaHOT}, AFM-SC spintronics~\cite{AKN_NatCommun19} in near future.


\emph{Acknowledgments.---} The first-principle calculations in this letter is supported by the Swedish National Infrastructure for Computing (SNIC) facility and for that, S.B. and A. K. N. sincerely thank Prof. P. M. Oppeneer.  P.C., S.B, A.K.G., A.S. and A. K. N. acknowledge the support from Department of Atomic Energy (DAE), Govt. of India. S B. and A.K.N. acknowledge the KALINGA HPC facility at NISER, Bhubaneswar and P.C., A.K.G. and A.S. acknowledge the SAMKHYA: HPC Facility provided at IOP, Bhubaneswar, for numerical computations. 
\bibliography{bibfile}{}

\begin{thebibliography}{99}
\bibitem [{spe()}]{spec_func}%
  \BibitemOpen
  \href@noop {} {}\bibinfo {note} {The spectral function
  $\mathcal{A}(k_x,\omega)$ has been defined through the zero-temperature
  Green's function $\hat{G}(k_x,\omega)$ as
  $\mathcal{A}(k_x,\omega)$=$-\frac{1}{\pi}{\rm Im}[{\rm
  Tr}\{\hat{G}(k_x,\omega)\}]$; where
  $\hat{G}(k_x,\omega)$=$\left[\mathbb{I}-\tilde{H}_{\rm L} (y,k_x)
  \right]^{-1}$}\BibitemShut {NoStop}%
\bibitem [{opt()}]{opt_lat}%
  \BibitemOpen
  \href@noop {} {}\bibinfo {note} {Within the generalized gardient apprximation
  (GGA) functional, the optimized lattice constant of bulk body-centered-cubic
  Nb is found to be $a_\textrm{Nb}$=3.3232~\AA, an excellent agreement with the
  experimental value of about 3.32~\AA~\cite{Nb_expt}. The value of
  $a_\textrm{Nb}$ is then used to constrauct the optimized film geometry and
  also strains are measured with respect to $a_\textrm{Nb}$.}\BibitemShut
  {Stop}%
\end{thebibliography}
\clearpage
\newpage

\begin{onecolumngrid}
	\begin{center}

		{
			\fontsize{12}{12}
			\selectfont
			\textbf{Supplemental material for ``Topological Superconductivity by Engineering Noncollinear Magnetism in
				Magnet/Superconductor Heterostructures: A Realistic Prescription for 2D Kitaev Model''\\[5mm]}
		}
		
		\normalsize Pritam Chatterjee\orcidA{}$^{1,2}$, Sayan Banik\orcidH{}$^{3}$, Sandip Bera\orcidD{}$^{3}$, Arnob Kumar Ghosh\orcidC{}$^{1,2}$, Saurabh Pradhan\orcidE{}$^{4}$, Arijit Saha\orcidF{}$^{1,2}$, and Ashis K. Nandy\orcidG{}$^{3}$\\
		{\small $^1$\textit{Institute of Physics, Sachivalaya Marg, Bhubaneswar-751005, India}\\[0.5mm]}
		{\small $^2$\textit{Homi Bhabha National Institute, Training School Complex, Anushakti Nagar, Mumbai 400094, India}\\[0.5mm]}
		{\small $^3$\textit{School of Physical Sciences, National Institute of Science Education and Research, An OCC of Homi Bhabha National Institute, Jatni 752050, India}\\[0.5mm]}
		{\small $^4$\textit{Lehrstuhl f\"ur Theoretische Physik II, Technische Universit\"at Dortmund Otto-Hahn-Str. 4, 44221 Dortmund, Germany}\\[0.5mm]}
	\end{center}	
		
\normalsize
\begin{center}
	\parbox{16cm}{}
\end{center}
\newcounter{defcounter}
\setcounter{defcounter}{0}
\setcounter{equation}{0}
\renewcommand{\theequation}{S\arabic{equation}}
\setcounter{figure}{0}
\renewcommand{\thefigure}{S\arabic{figure}}
\setcounter{page}{1}
\pagenumbering{roman}
\setcounter{section}{0}
\renewcommand{\thesection}{S\arabic{section}}

\vspace{-1cm}
\section{Topological characterization} \label{Sec:S1}
In this section of the supplementary material (SM), we discuss the topological characterization of our model. We begin with the continuum Hamiltonian described in the main text [Eq.~(3)]. One can obtain the lattice version of the continuum Hamiltonian by replacing $k_x\to\sin k_x$, $k_y\to\sin k_y$, $(1-k_x^2/2)\to\cos k_x$, and $(1-k_y^2/2)\to\cos k_y$ as 
\begin{align}
	\tilde{H}_{\rm L}(\vect{k})=&\xi_{\vect{k},\vect{g}}\tau_{z}+\frac{1}{2}(g_x\sin k_x+g_y\sin k_y)\sigma_z\tau_{z}+J\sigma_x+\Delta_{0}\tau_{x} \ , 
	\label{fullmodifiedhamiltonian}
\end{align}
where, $\xi_{\vect{k},\vect{g}}= (2-\cos k_x-\cos k_y)+\frac{(g_x^2+g_y^2)}{2}-\mu$. At the high symmetry points $\vect{\Gamma}_l=[(0,0),(\pi,0),(0,\pi),(\pi,\pi)]$, if we perform the proper basis transformation, then the Hamiltonian in Eq.~(\ref{fullmodifiedhamiltonian}) can be recast in the block diagonal form as,
\vskip -0.8cm
\begin{eqnarray}
	\tilde{H}_{\rm L}(\vect{k}) &= \begin{pmatrix}
		h(\vect{\Gamma}_l) & O_{2\times 2}
		\\
		O_{2\times 2}                           &
		-h(\vect{\Gamma}_l) 
	\end{pmatrix} \ .
	\label{Matrix1}
\end{eqnarray}
where,
\begin{eqnarray}
	h(\vect{\Gamma}_l) &= \begin{pmatrix}
		J+\xi_{\vect{k},\vect{g}}(\vect{\Gamma}_l) & \Delta_{0}
		\\
		\Delta_{0}                           &
		J-\xi_{k,g}(\vect{\Gamma}_l) 
	\end{pmatrix} \ .
	\label{Matrix2}
\end{eqnarray}
\begin{figure}[H]
	\begin{center}
		\includegraphics[width=0.48\textwidth]{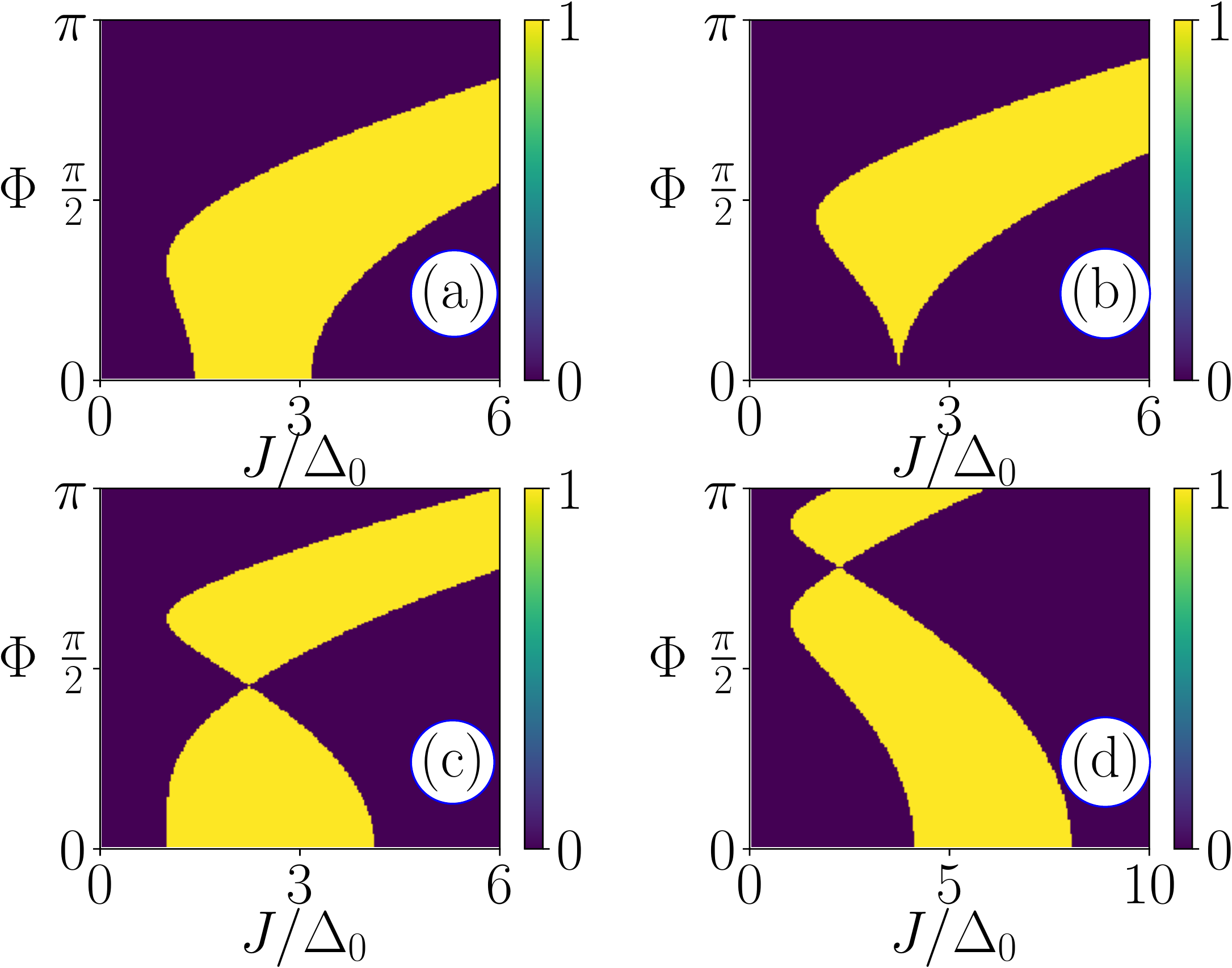}
	\end{center}
	\caption{Topological invariant $\nu$ for the symmetric ($g_x=g_y=\Phi$) spin texture is shown. 
		Panels (a), (b), (c), and (d) correspond to the variation of $\nu$ in the $J \mhyphen \Phi$ plane for different values of the chemical potential $\mu=\Delta_0, 2.0\Delta_0, 4.0\Delta_0$, and 
		$8.0\Delta_0$, respectively.} 
	\label{Fig6}
\end{figure}

\begin{figure}[H]
	\begin{center}
		\includegraphics[width=0.48\textwidth]{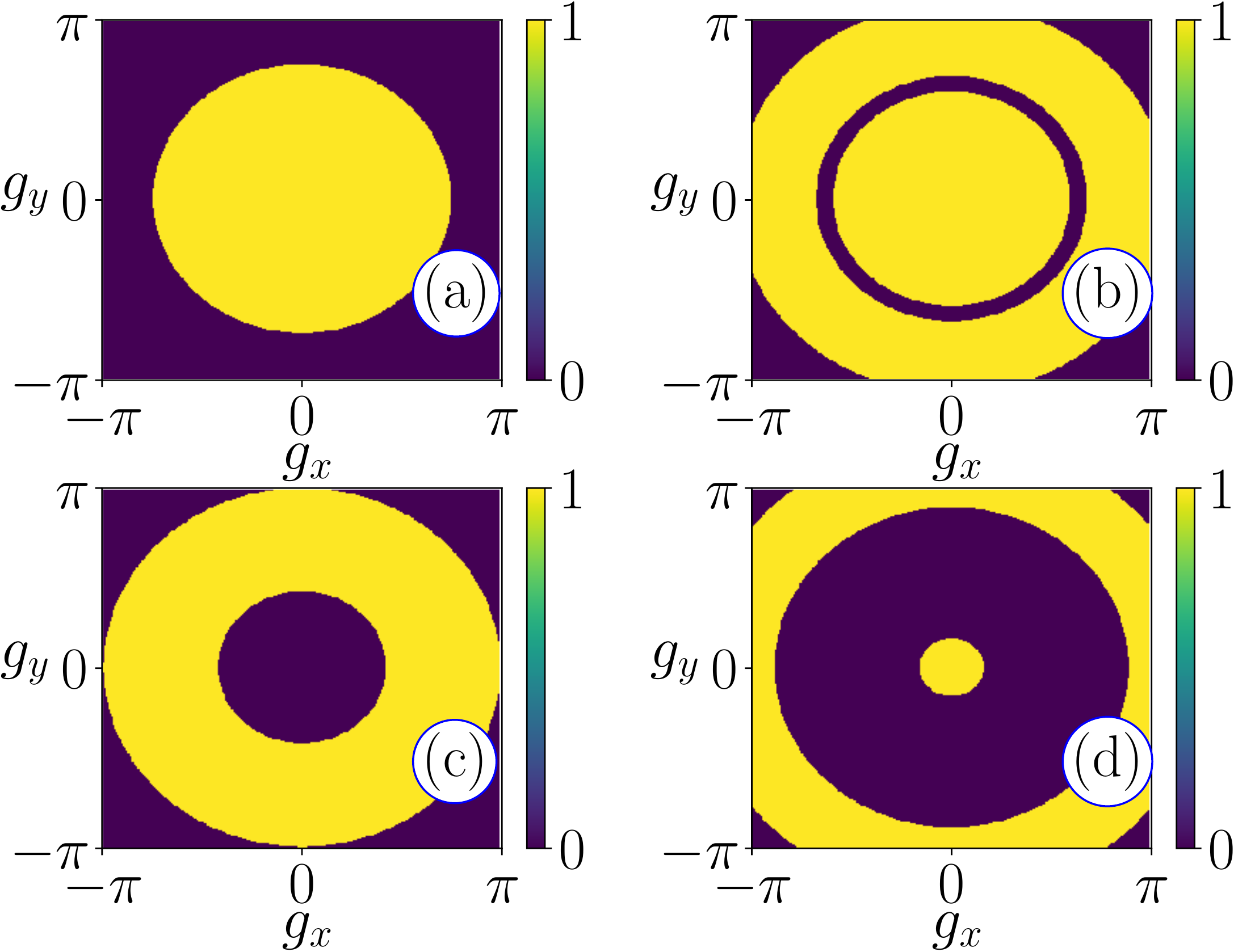}
	\end{center}
	\caption{\hspace*{-0.1cm}Topological invarient $\nu$ in case of asymmetric spin texture ($g_x\ne g_y$) is depicted.  Panels (a), (b), (c), (d) correspond to the variation of $\nu$ in $g_x-g_y$ plane, where we choose the other parameter values for panels (a), (b) are $J=2.0\Delta_0, \mu=\Delta_0$ and $J=2.0\Delta_0,\mu=4.0\Delta_0$ respectively. On the other hand, for panels (c), (d) we choose $J=4.0\Delta_0, \mu=\Delta_0$ and $J=4.0\Delta_0, \mu=4.0\Delta_0$ respectively.} 
	\label{Fig7}
\end{figure}
The topological invariant reads~\cite{Bena2015},
\begin{eqnarray}
	\delta &= {\rm sgn}~\prod_{l=1}^4\det\begin{pmatrix}
		J+\xi_{\vect{k},\vect{g}}(\vect{\Gamma}_l) & \Delta_0\\
		\Delta_0&J-\xi_{\vect{k},\vect{g}}(\vect{\Gamma}_l)
	\end{pmatrix}\ 
	&=(-1)^\nu\ ,
\end{eqnarray}

where the quantity $\nu$ is treated as a topological invariant. Therefore, one can identify the topologically trivial and non-trivial regime by considering $\nu=0$ or $\nu=1$ accordingly. In Fig.(\ref{Fig6}) 
and Fig.(\ref{Fig7}) we illustrate the topological invariant considering symmetric and asymmetric spin spirals (SS) in different parameter regimes. In each figure (Figs.~\ref{Fig6} and \ref{Fig7}), the yellow region indicates the gapless topological superconducting (TSC) phase ($\nu=1$) hosting Majorana flat edge modes (MFEM) due to the generation of an effective $(p_x+p_y)$ type of superconducting pairing. On the other hand, in the violet region, the system becomes a topologically trivial gapped superconductor ($\nu=0$), where MFEM disappear due to the destruction of the $(p_x+p_y)$ type pairing.

\section{Methodology}\label{Sec:S2}
In this section, we discuss the computational details for extracting the material specific parameters within \textit{ab~initio} electronic structure calculations. All these parameters are for various heterostructures made of an atomic-layer-thick magnetic layer of 3$d$ transition metal (TM) elements, Mn and Cr, on the top of normal $s$-wave superconducting substrate. A body-centred-cubic ($bcc$) Nb with different surface cuts, (110) and (001) is used as the substrate. Ultimately, these parameters are used for generating ground state spin texture by solving extended Heisenberg model within Monte Carlo (MC) simulations~\cite{MC_1,MC_2}, which is the essential ingredient for establishing the topological superconducting (TSC) phase.
\subsection{Film geometry relaxation within \textit{ab~initio} electronic structure method}
We first begin with the optimized lattice of bulk $bcc$-Nb superconductor. The optimized lattice constant of bulk $bcc$-Nb has been obtained by performing first-principles electronic structure calculations within the density-functional theory and the plane-wave pseudopotential approach as implemented in the Vienna Ab-initio Simulation Package (VASP)\cite{hafner2008ab,kresse1996,Kresse}. Exchange and correlation have been considered within the generalized gradient approximation (GGA) which is parametrized via Perdew, Burke, and Ernzerhof (PBE) functional~\cite{Perdew1996}. Here, the projector-augmented wave (PAW) method is employed to construct the pseudopotentials with valence wavefunctions which are approximated smooth near ion cores~\cite{paw1, paw2}. The plane wave cut-off energy (500~$\rm eV$) and the $k$-point mesh (16$\times$16$\times$16 and $\Gamma$-centred) in the full Brillouin-zone (BZ) are checked carefully for the BZ integration, ensuring the numerical convergence of self-consistently determined quantities. The optimised lattice constant of $bcc$-Nb is found to be $a_\textrm{opt}$=3.3232 \AA, matching well with the experimental value in the Materials Project database ($a_\textrm{expt}$=3.32 \AA)~\cite{Nb_expt}.

The surface unit cell (u.c.) in the form of two-dimensional (2D) slab geometry is constructed with sufficiently thick vacuum layers. The thickness of vacuum on each side of the slab is taken about 10~\AA. Assuming a pseudomorphic layer by layer growth, ultrathin magnetic films are constructed by considering a single layer of Mn and Cr on a well known $s$-wave superconductor (SC) substrates Nb(110) and Nb(001), respectively.  In case of Mn/Nb(110) film, each surface layer contains two atoms and the thickness of the substrate is taken 11 monolayers (MLs),~see Fig.~\ref{Mn_Nb110}(a). In the Cr based film as depicted in Fig.~\ref{S5.fig}(a), one atom per layer is repeated along $c$-axis with the substrate thickness of about 15 MLs. The surface u.c. parameters using optimized $bcc$-Nb lattice constants are $a$= 3.3232~\AA~with $b/a$=$\sqrt{2}$ and $a$=$b$=3.3232~\AA~for Mn/Nb(110) and Cr/Nb(001) magnetic films, respectively. Increasing the layer numbers beyond that doesn't carry any effect on the properties of the systems. Later, uniform biaxial strain has been calculated with respect to the optimized lattice constant as, strain=$\frac{a-a_\textrm{opt}}{a_\textrm{opt}}\!\times\!100 \%$. Therefore, the sign of strain in case of expansion (compression) is positive (negative). The number of MLs in the substrate and the vacuum thickness for all film structures with both compressive and tensile strains are kept fixed. We have used the same energy cut-off for the plane waves and the 12$\times$8$\times$2 $k$-mesh for the BZ integration. The distances between different layers counting from the top TM layer till the 8$^{\rm{th}}$ layer (7 interlayer distances) in all films are relaxed until the Hellmann-Feynman force on each atom is smaller than 0.001~$\rm {eV}$/\AA. So, we have applied both positive and negative values of strain ($< 5 \%$) on both Mn/Nb(110) and Cr/Nb(001) films and relaxed each of them along the growth direction.  We find that further relaxation with more layers ($>$ 8) does not change the computed results significantly. In every calculations, the self-consistent energy has been converged with an accuracy of 10$^{-7}~\rm {eV}$.

In Table~\ref{tab01} and \ref{tab02}, we have provided the theoretical relaxed structure for Mn/Nb(110) and Cr/Nb(001) magnetic films, respectively. The top Mn and Cr layers are assigned with the number $n$=0 in Fig.~\ref{Mn_Nb110}(a) and \ref{S5.fig}(a), respectively and $n$ increases if one moves down across the film. The  relaxed interlayer distances till $n$=4 (first four interlayer separations only) are presented in those Tables and in the parentheses, we provide the changes in \% with respect to the ideal interlayer spacing. For example, the ideal interlayer distance in case of Nb(110) and Nb(001) films using the optimized $bcc$-Nb lattice are 2.3499 \AA~and 1.6616 \AA, respectively. It is expected that the changes in interlayer distance (in \%) will be more with both negative and positive strains. Note, we have carefully checked the optimum numbers (till $n=6$) of relaxed interlayer distances, beyond that the change in the interested parameters is negligible.

\begin{table}[H]
	\centering
	\begin{tabular}{ |c|c|c|c|c|c| } 
		\hline
		Lattice constant  & $d_{01}$ in~\AA  &$d_{12}$ in~\AA &$d_{23}$ in~\AA &$d_{34}$ in~\AA  \\ 
		in \AA ($a_\textrm{strain}$)&(\% change)&(\% change)&(\% change)&(\% change)\\ \hline
		3.432 ($a_{+3.3}$)&1.97(+18.8\%)&2.24(+7.7\%)&2.24( +7.7\%)&2.24( +7.7\%)\\ \hline
		3.366 ($a_{+1.3}$)&2.00(+16.0\%)&2.31(+3.1\%)&2.29(+11.4\%)&2.30(+11.1\%)\\ \hline
		3.3232 ($a_{0.0}$)&2.03(+13.5\%)&2.35(+0.2\%)&2.34( +0.3\%)&2.35( +0.2\%)\\ \hline
		3.270($a_{-1.6}$)&2.08(+10.0\%)&2.40($-$3.8\%)&2.42( $-$4.7\%)&2.42( $-$4.7\%)\\ \hline
		3.234 ($a_{-2.7}$)&2.10( +8.2\%)&2.44($-$6.6\%)&2.45( $-$7.0\%)&2.44( $-$6.6\%)\\ \hline
		3.200($a_{-3.7}$)&2.14( +5.4\%)&2.47($-$9.1\%)&2.48($ -$9.5\%)&2.48( $-$9.5\%)\\ \hline
	\end{tabular}
	\caption{Structural relaxation details for Mn/Nb(110) magnetic film. Here, $d_{nm}$ defines the distance between layers after relaxation. The `+ve' (`$-$ve') signs in $d$ signifies that relaxed distance is larger (smaller) than the ideal distance in the film, obtained from $a$ in each case.}  
	\label{tab01}
\end{table}

\begin{table}[H]
	\centering
	\begin{tabular}{ |c|c|c|c|c|c| } 
		\hline
		Lattice constant  & $d_{01}$ in~\AA  &$d_{12}$ in~\AA &$d_{23}$ in~\AA &$d_{34}$ in~\AA  \\ 
		in \AA ($a_\textrm{strain}$)&(\% change)&(\% change)&(\% change)&(\% change)\\ \hline
		3.432 ($a_{+3.3}$)&1.22(+28.9\%)&1.65(+3.6\%)&1.55(+9.8\%)&1.6(+6.7\%)\\ \hline
		3.366 ($a_{+1.3}$)&1.29(+23.4\%)&1.67(+0.9\%)&1.6(+5.4\%)&3.65(+1\%)\\ \hline
		3.3232($a_{0.0}$)&1.31(+21.07\%)&1.68(-1.3\%)&1.61(+3.1\%)&1.66(0.0\%)\\ \hline
		3.300(a$_{-0.7}$)&1.30(+21.3\%)&1.72(-4.2\%)&1.63(+0.1\%)&1.67(-1.6\%)\\ \hline
		3.234 ($a_{-2.7}$)&1.36(+15.89\%)&1.74(-7.5\%)&1.69(-4.6\%)&1.71(5.7\%)\\ \hline
		3.168 ($a_{-4.7}$)&1.38(+12.88\%)&1.78(-12.5\%)&1.75(-10.6\%)&1.73(-9.5\%)\\ \hline
	\end{tabular}
	\caption{Structural relaxation details for Cr/Nb(110) magnetic film. Here, $d_{nm}$ defines the distance between layers after relaxation. The `+ve' (`$-$ve') signs in $d$ signifies that relaxed distance is larger (smaller) than the ideal distance in the film, obtained from $a$ in each case.}
	\label{tab02}
\end{table}

\subsection{Extraction of material specific parameters: Korringa-Kohn-Rostokar Green-function (KKR-GF) method} 
After the structural relaxation of each magnetic film, the relaxed structures are now used for \textit{ab~initio} simulations performed within the scalar-relativistic screened KKR-GF method~\cite{Bauer_thesis} with exact description of the atomic cells. This method is based on the multiple-scattering theory~\cite{sp_kkr} that consists of dividing the problem in calculating the electronic structure of a solid into two parts: solving a single-site scattering problem for each atom in isolation, then incorporating the structural information of the solid by solving a multiple-scattering problem. Note, all the magnetic films are enclosed by two vacuum regions with a thickness of about 10~\AA~on both top and bottom sides. 

The local spin density approximation (LSDA) has been considered~\cite{vwn80}. The effective potentials and fields are treated within the atomic sphere approximation (ASA) with an angular momentum cut-off, $l_\textrm{max}$=3. The energy integrations are performed using a grid of 38 points along a path of the complex-energy contour with a Fermi smearing value of 473 K. For the necessary $k$ integrations in the 2D BZ, we have chosen 1600 (40×40×1) $k$-points in the full surface BZ for the integration of the Matsubara pole closest to the real axis in order to perform the self-consistency. Within KKR-GF method including spin-orbit coupling, we have first converged the potential self-consistently for each film, with confining the magnetic moments along out-of-plane direction ($z$-direction).

By employing the KKR-GF method, we essentially calculate the following important parameters: the symmetric exchange interaction, $\mathcal{J}_{ij}$'s, the antisymmetric Dzyaloshinskii-Moriya interaction (DMI), $\textbf{D}_{ij}$ ($\vert \textbf{D}_{ij}\vert$ = $\mathcal{D}_{ij}$), and  the single ion magnetocrystalline anisotropy (MCA), $\mathcal{K}_i$. These parameters contribute in the following Heisenberg model Hamiltonian described in Eq.~(\ref{spin_hamiltonian}) and finally, the numerical solutions within MC simulations describe the magnetic textures in the magnetic layer.

\vspace{-0.25cm}
\begin{align}
	\mathcal{H}&=-\sum_{i>j}\big[\!\mathcal{J}_{ij}\hat{n}_i.\hat{n}_j+\textbf{D}_{ij} \cdot (\hat{n}_i\times\hat{n}_j)\big]-\mathcal{K}\sum_i(\hat{n}_i\cdot \hat{z})^2 \ ,
	\label{spin_hamiltonian}
\end{align} 
where $i$ and $j$ denote the site of atoms in a considered domain and the corresponding $\hat{n}_i$ and $\hat{n}_j$ represent unit vectors along the magnetic moments. In this model, the negative (positive) sign of $\mathcal{J}$ signifies antiferromagnetic (ferromagnetic) pair-wise coupling. This spin Hamiltonian is also describes various topological magnetic states in the presence of external perturbation~\cite{AKN_ScRep,AKN_PRB17}.

The advantage of KKR-GF method is that one can calculate two-site exchange interactions, both isotropic $\mathcal{J}$ and anisotropic $\textbf{D}$ vector. Here, the non-zero value of DMI with its orientation actually decides the nature of the chiral spin-spiral (SS). Once the converged potential is obtained self-consistently, we perform three single-shot calculations maintaining magnetization along the $x, y$ and $z$ directions and employing the infinitesimal rotations method~\cite{LIECHTENSTEIN198765} as implemented within a relativistic generalized formalism~\cite{Noncol1,NonCol2}. These allow us to determine all three components of the DMI vector.
We took the cutoff radius of seven in unit of the lattice constant (=7$a$ \AA) to capture the long-range interactions found in the system. This includes more than 25 intralayer shells for which two-site parameters are extracted. In order to calculate the MCA, the converged potential is further used to calculate the total band energy via one-shot calculations for magnetization oriented along three orthogonal directions ($x, y$ and $z$). Here, a larger $k$-point mesh (80$\times$80$\times$1=6400) is considered. With this consideration, we obtain the MCA as,

\vspace{-0.25cm}
\begin{align} 
	\mathcal{K}_x=E_x-E_z ~\textrm{and}~
	\mathcal{K}_y=E_y-E_z \ ,
\end{align}
where, positive values of $\mathcal{K}_x$ and $\mathcal{K}_y$ clearly refer to the out-of-plane anisotropy and here, the minimum value between them refers to the single ion MCA constant, $\mathcal{K}$. The negative value of $\mathcal{K}$ on the other hand refers to the in-plane anisotropy.
\vskip +0.4cm

\subsection{Numerical calculations for finding the magnetic ground state: Atomistic Spin Dynamics simulations}
After extracting all magnetic interaction parameters required for the ground state magnetic textures of Mn/Nb(110) and Cr/Nb(001) systems, we have numerically solved the Eq.~\ref{spin_hamiltonian} within Monte Carlo (MC) simulations~\cite{MC_1,MC_2,Bera2019,Bera2021}. This will give the best atomistic description of our magnetic systems within the Atomistic Spin Dynamics simulations code $Spirit$~\cite{SPIRIT}. The ground state spin textures are found to be robust by means of no change in the ground state configuration with more interaction shells. We performed the simulated annealing process~\cite{MC_2} to identify the zero-temperature ground state of the system. In such process, we have considered 10$^6$ MC steps followed by 10$^4$ thermalization steps at each temperature step. We started from a finite temperature in the range between 15K and 30 K and reached the zero-temperature ground state configuration with atleast $\times$10$^2$ number of steps. A typical size of the simulation domain in our MC simulations is fixed to 32$\times$32, after carefully checking the ground state configuration in a larger domain of size like 128$\times$128.

\section{Example of $\textrm{Mn/Nb}$(110) film: optimized and uniformly planar strained structures}

\subsection{Mn/Nb(110) film with optimized lattice parameter of $bcc$-Nb}\label{Sec:S3}
In this subsection of the SM, we first consider a Mn/Nb(110) film in Fig~\ref{Mn_Nb110}(a) constructed with the GGA-optimized lattice parameter of $bcc$-Nb, forming a $c(2\times2)$ surface unitcell with lattice constants, $a_\textrm{opt}$=3.3232 \AA~and $b_\textrm{opt}$=$\sqrt{2}a_\textrm{opt}$=4.6997 \AA. After relaxation, the calculated exchange parameters, $\mathcal{J}$ and $\mathcal{D}$, are depicted in Fig.~\ref{Mn_Nb110}(b) as a function of the distance measured in the unit of the smallest surface lattice parameter \ie $a_\textrm{opt}$. The real distance from an atom at position $\vect{R_0}$ to the other atom at $l^{\rm{th}}$ neighboring shell can be defined as $\vert\vect{R}_l-\vect{R}_0\vert$. A strong frustration in $\mathcal{J}$'s is observed in Fig.~\ref{Mn_Nb110}(b) and as a result, we find a $c(2\times 2)$ antiferromagnet (AFM) as the ground state in our MC simulations, see Fig~\ref{Mn_Nb110}(c). The out-of-plane anisotropy constant is found to be small. A significantly weak DMI strength (the ratio between strongest $\mathcal{D}$ and $\mathcal{J}$ is $\approx$ 0.02) cannot support any noncollinearity in the $c(2\times2)$-AFM structure. The nearest-neighbor (NN) AFM exchange coupling and the next-nearest-neighbor (NNN) ferromagnetic (FM) exchange coupling with an optimal ratio may support such $c(2\times2)$-AFM phase as the ground state. Indeed, the ground state of Mn/Nb(110) reported in a recent experiment by Conte \etal~\cite{Conte2022} have now been well reproduced within our theoretical model. The simulated spin texture in Fig~\ref{Mn_Nb110}(c) clearly manifests an AFM order along [110] and [$\bar{1}$10] directions and FM order along [100] and [010] directions. 

\begin{figure}[H]
	\begin{center}
		\includegraphics[width=1\textwidth]{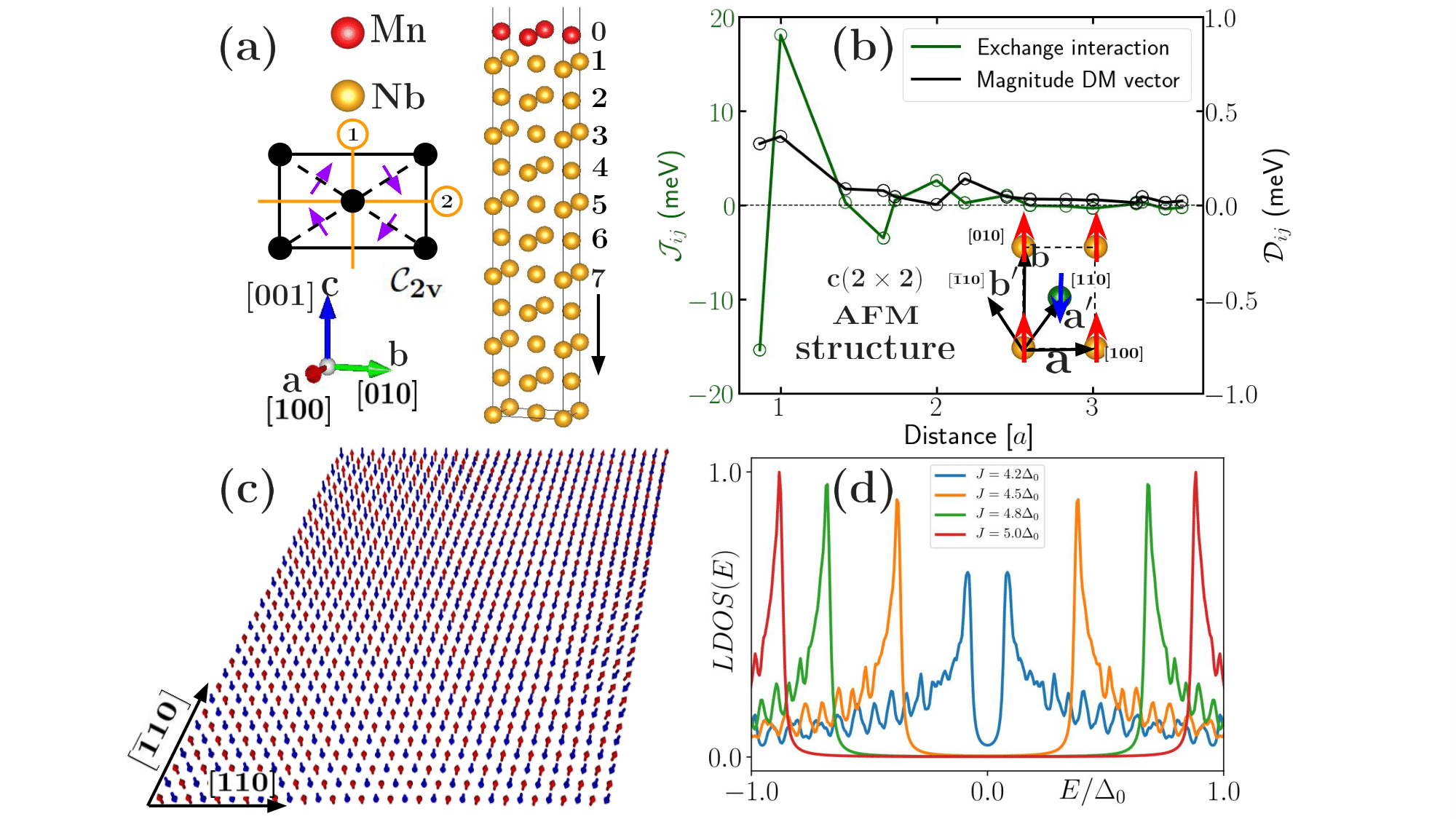}
	\end{center}
	\caption{\hspace*{-0.1cm} In panel (a), the unit cell of Mn/Nb(110) used for our \textit{ab~initio} calculations. The atoms denoted by red and yellow balls are Mn and Nb atoms, respectively. On the left, the sketch of the rectangular $c(2\times2)$ surface u.c. has $\mathcal{C}_{2v}$ symmetry with mirror planes ($p$=1, 2)  are indicated by yellow lines. The filled black circles represent the surface atoms in the u.c. and the microscopic $\vect{D}$ vectors (purple arrows) between NN sites on crystal surface plane are restricted by symmetry to point perpendicular to the bond. Indeed, the calculated vector orientations of all $\vect{D}$ vectors in the inset of Fig.~4(a) in the main text follow the $\mathcal{C}_{2v}$ symmetry rules~\cite{DMI2}. In panel (b), we depict the exchange coupling strength, $\mathcal{J}_{ij}$ and DMI strength, $\mathcal{D}_{ij}$ as a function of distance from a reference atom. The interaction parameters, $\mathcal{J}$'s and $\mathcal{D}$'s, till 5$^{\rm{th}}$ neighbor cell is also presented in Table~\ref{tab1}. We find the expected decay of both $\mathcal{J}$ and $\mathcal{D}$ with distance $\vert\vect{R}_l-\vect{R}_0\vert$ in units of surface lattice vector $a_\textrm{opt}$. In panel (c), we show an AFM spin texture which is found in the Mn layer within the MC simulation. We have used parameters mentioned in (b) and out-of-plane anisotropy, $\mathcal{K}$ = 0.04 $\rm {meV}$ per Mn atom. Panel (d) shows the plot of the local density of states (LDOS) as a function of energy, illustrating the formation of Shiba bands within $\pm\Delta_0$ for the $c(2\times2)$-AFM phase (see the inset of (b)). The exchange field strength $J$ is varied between 4.2$\Delta_{0}$ and 5.0$\Delta_{0}$. We choose $\mu=4.0\Delta_0, t=\Delta_0$.}
	\label{Mn_Nb110}
\end{figure}

A domain of size 32$\times$32 has been used for solving Eq.~(4) in the main text and the obtained results are summarized in Fig.~\ref{Mn_Nb110}(d), describing the local density of states~(LDOS) as a function of energy ($E$). Due to the overlapping of the Yu-Shiba-Rusinov (YSR) in-gap bound states, the Shiba bands are formed within the superconducting gap $\pm\Delta_0$. We further vary the coupling constant $J$ between the local magnetization (in the Mn spin texture) and the itinerant spin (free electrons in the SC) to ensure the trivial superconductivity in the system. Precisely, a proximity induced SC sustains in the AFM Mn layer. Based on our results from our model (Eq.~(4) in the main text), we can firmly reproduce the experimental findings \ie the coexistence of SC and AFM phases~\cite{Conte2022}.

\subsection{Role of uniform biaxial surface strain for AFM-SS ground state in Mn/Nb(110) film}\label{Sec:S4}
Here, we discuss a transition to an AFM-SS state from the $c(2\times2)$ state in the top Mn layer of Mn/Nb(110) heterostructure by applying an uniform biaxial strain in the 2D film geometry. As the strain changes the in-plane lattice parameters keeping $b$=$\sqrt{a}$ unaltered, the relaxed film is expected to change $\mathcal{J}$'s, $\mathcal{D}$'s and $\mathcal{K}$ value for the magnetic layer. In Table~\ref{tab1}, we have presented various quantities calculated within KKR-GF method while the Fig.~\ref{S4.fig}(a) depicts the effect of strains on $\mathcal{J}$'s as a function of distance between Mn atoms. In the second column of the Table~\ref{tab1}, the magnetic moment of Mn within KKR-GF calculation is matching well with that of obtained from VASP relaxation calculations. The plot in Fig.~\ref{S4.fig}(a) ensures the strong frustration in $\mathcal{J}$'s even after varying the planar strain 
from tensile (positive) to compressive (negative). Indeed, the dominating parameters are the Heisenberg exchange parameters and the first two $\mathcal{J}$'s are the strongest in comparison to the rest. In particular, the NN exchange constant (`$-$ve' means AFM) is gradually increasing in magnitude with changing strain from tensile to compressive while the trend is opposite in case of NNN (`$+$ve' means FM) exchange constant except for $a_{+3.3}$ case, see Table~\ref{tab1}.  

Interestingly, this strong exchange frustration has adverse effect in the ground state magnetic phase of Mn/Nb(110), although, the $c(2\times2)$-AFM phase remains the lowest energy magnetic configuration under the tensile strain \ie when the 2D surface u.c. is expanded. An AFM-SS state becomes the lowest energy ground state under a small compressive strain of magnitude 1.6 \%. It is worth to mention that such AFM-SS ground state is rarely found in film geometry and the number of examples is very limited~\cite{Bode_Nat07}. Here, this SS state in Fig.~\ref{S4.fig}(c) can be stabilized by the exchange frustration even in the presence of small out-of-plane $\mathcal{K}$. We further increases the compressive strain and the AFM-SS solution is found with a small variation in the period $\lambda_\textrm{Mn}$ for strains $-$2.7\% and $-$3.7\% both. We find that the AFM-SS state exists for a delicate balance between NN AFM and NNN FM $\mathcal{J}$ values, hardly affected by the rest of isotropic exchange interaction terms. The spin spiral presented in Fig.~\ref{S4.fig}(c) for $-$2.7 \% strain has a period of about $\lambda_\textrm{Mn} \approx$ 2.82 $\rm nm$ computed using MC simulations. In such cases, the sense of rotation of the SS will have degeneracy which generally breaks in the presence of chiral interactions \ie the DMI. Hence, using the full Hamiltonian in Eq.~(\ref{spin_hamiltonian}), we have identified a right-handed cycloidal AFM-SS state, see Fig.~\ref{S4.fig}(b) and also Fig.~4(b) in the main text. Comparing (b) and (c), we find that even relatively weak DMI strength changes the period of the SS to an expected lower value, $\lambda_\textrm{Mn} \approx$ 2.35 $\rm nm$, propagating along [010] direction.  In contrast, the AFM-SS earlier reported in Mn/W(110), a non-SC film system, where a strong DMI determines the left-rotating spin cycloid, owing to the strong SOC from the heavy metal element W~\cite{Bode_Nat07}.

The exchange frustration driven AFM-SS is now examined by numerically solving Eq.~(4) in the main text. A 32$\times$32 domain as presented in Fig.~\ref{S4.fig}(d) is considered for the spin-lattice model and the corresponding result, particularly the LDOS for the zero-energy ($E=0$) is depicted in Fig.~\ref{S4.fig}(e). The zero-energy states are indeed localized at the edges of the considered magnetic domain while the bulk YSR band is semimetallic. Therefore, this AFM-SS can give rise to MFEM mode (see Fig.~\ref{S4.fig}(e)), even without DMI. 
Akin to the main text (see inset of Fig.~5(a)), non-dispersive states at the zero-energy is also observed in the eigenvalue spectrum. In general, the antisymmetric DMI term in magnetic films with broken inversion symmetry is the result of an indirect exchange mechanism in the presence of spin-orbit coupling (SOC) present in the substrate elements~\cite{DMI1,DMI2,FertPRL80}. In many 2D magnetic systems, this plays an important role in stabilizing SS state which otherwise does not appear with the exchange interactions only~\cite{NandyNanoLett20}. This is also true in case of Mn/Nb(110) with optimized lattice constant, see the subsection~\ref{Sec:S3}. Hence, for Mn/Nb(110) magnetic film, strain is an important controlling parameter to stabilize the AFM-SS state and hence, triggers the TSC phase. Comparing the results presented in the main text Fig.~5(a)-(b) and Fig.~\ref{S4.fig}(e), we have established that the TSC phase transition from a trivial state can occur when the noncollinear spin textures in the form of SS becomes a stable solution in a transition-metal/superconductor (TM/SC) heterostructure even in the absence of DMI, a SOC driven interaction parameter. 

\begin{table}[H]
	\centering
	\begin{tabular}{|c|c|c|c|c|c|c|c|c|c|c|c|c|}\hline
		Lattice&Magnetic mom.&MCA& \multicolumn{4}{|c|}{DMI magnitude}&\multicolumn{5}{|c|}{Isotropic exchange}&Ground  state \\ 
		constant in \AA&$\mathcal{M}_\textrm{Mn}$ in $\mu_\textrm{B}$&$\mathcal{K}$ in & \multicolumn{4}{|c|}{in $\rm{meV}$}&\multicolumn{5}{|c|}{interaction ($\rm{meV}$)}&spin texture  \\ \cline{4-12} ($a_\textrm{strain}$)&KKR-GF,~(VASP)&${\rm{meV}}$/Mn&$\mathcal{D}_{01}$&$\mathcal{D}_{02}$&$\mathcal{D}_{03}$&$\mathcal{D}_{04}$&$\mathcal{J}_1$&$\mathcal{J}_2$&$\mathcal{J}_3$&$\mathcal{J}_4$&$\mathcal{J}_5$&\\ \hline
		3.432 ($a_{+3.3}$)&3.63, (3.51)&0.22&0.53&0.5&0.02&0.05&-12.41&17.96&2.78&-2.6&1.25&AFM\\ \hline
		3.366 ($a_{+1.3}$)&3.57, (3.46)&0.17&0.42&0.52&0.04&0.07&-14.44&18.84&1.34&-3.18&0.78&AFM\\ \hline
		3.3232 ($a_{0.0}$)&3.56, (3.43)&0.04&0.33&0.37&0.09&0.08&-15.38&18.01&0.03&-3.49&0.5&AFM\\ \hline   
		3.270 ($a_{-1.6}$)&3.55, (3.43)&0.22&0.28&0.19&0.14&0.07&-16.58&16.16&-0.57&-3.38&0.15&AFM-SS\\ \hline
		3.234 ($a_{-2.7}$)&3.53, (3.39)&0.22&0.28&0.11&0.2&0.07&-17.89&15.37&-0.86&-3.26&0.17&AFM-SS\\ \hline     
		3.200 ($a_{-3.7}$)&3.52, (3.40)&0.29&0.28&0.04&0.27&0.05&-19.61&14.49&-0.67&-2.96&0.3&AFM-SS\\ \hline
	\end{tabular}
	\caption{The first three columns provide the smallest lattice constants under strain, $\mathcal{M}_\textrm{Mn}$ obtained from both KKR-GF and VASP for each Mn atom and single ion MCA constant (all are out-of-plane), respectively. Next, the $\mathcal{D}_{0j}$ and $\mathcal{J}_i$ for first few neighboring cells are tabulated. The different lattice constants signify the system is constructed with different uniform biaxial strain values: from top to bottom, tensile 3.3\% (a$_{+3.3}$), tensile 1.3\% (a$_{+1.3}$), optimized 0.0\% (a$_{0.0}$),  compressive 1.6\% (a$_{-1.6}$), compressive 2.7\% (a$_{-2.7}$),  and compressive 3.7\% (a$_{-3.7}$). The strains are calculated with respect to the optimised lattice constant, $a_\textrm{opt}$=3.3232 \AA. The last column clearly indicates a transition from a collinear AFM state to a noncollinear AFM-SS state under a small compressive strain.}
	\label{tab1}
\end{table}

\begin{figure}[H]
	\begin{center}
		\includegraphics[width=0.9\textwidth]{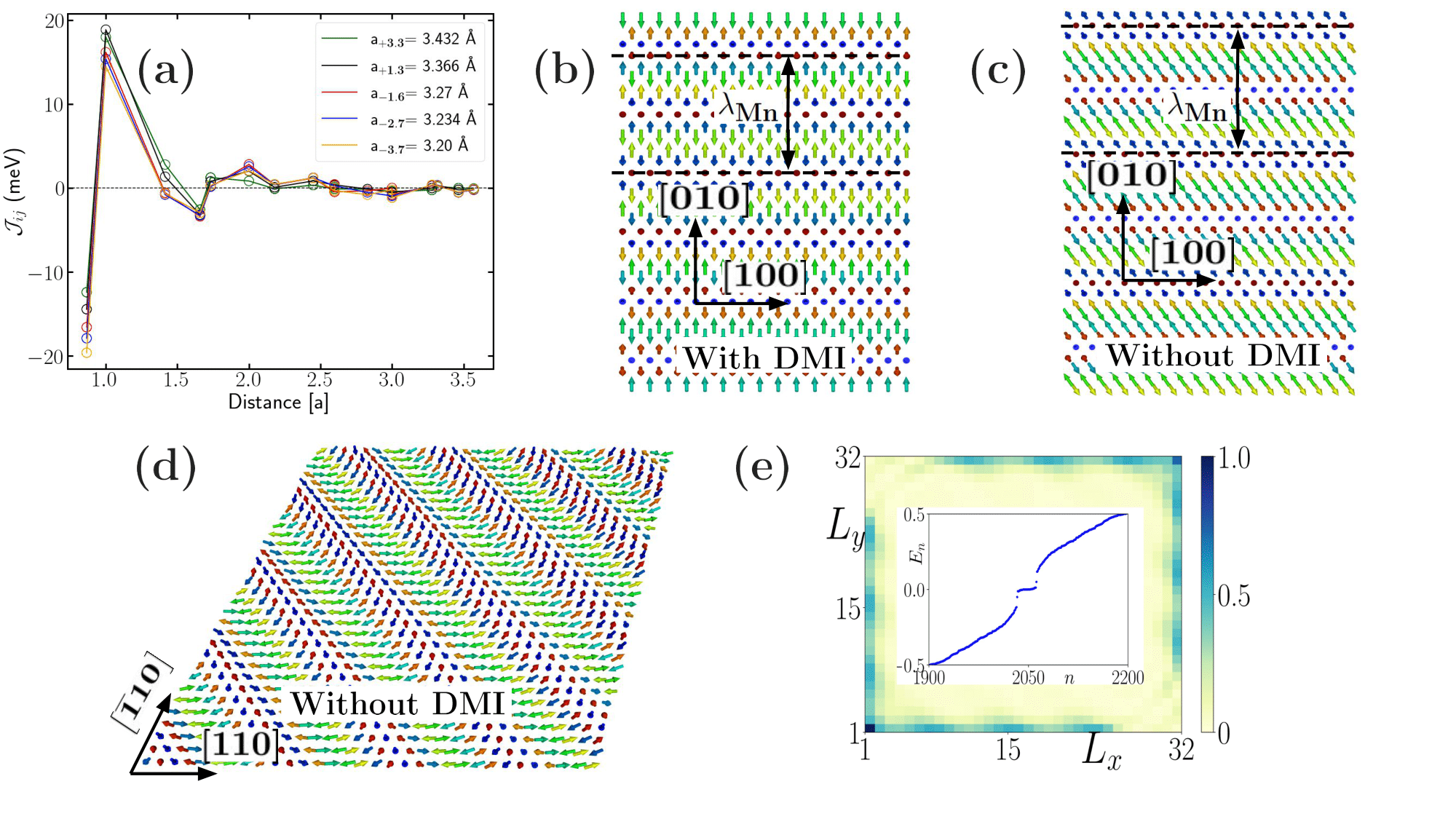}
	\end{center}
	\caption{\hspace*{-0.1cm} For various strain values, in panel (a) we depict $\mathcal{J}$'s as a function of distance. Spin textures in panels (b) and (c) illustrate the ground state AFM-SS solutions with DMI and without DMI, respectively. The magnitude of the compressive strain is fixed to 1.6. The AFM-SS state becomes an homogeneous right-handed AFM cycloid in the presence of DMI, propagating along [010] direction and the spins are rotated in the $yz$-plane. (d) An AFM-SS domain is constructed from the spin texture shown in panel (c) and hence edges of the domain are [110] and [$\bar{1}$10]. The normalized LDOS corresponding to $E$=0 eigenstate is shown in panel (d) in the $L_x \mhyphen L_y$ square plane. The inset depicts the corresponding eigenvalue spectrum $E_n$ as a function of the state index $n$. Hence, clearly we obtain TSC phase hosting MFEM in the absence of DMI. Here, we choose 
		$\mu=4.0\Delta_0,t=\Delta_0$ and $J=1.5\Delta_0$.} 
	\label{S4.fig}
\end{figure}
\section{Example of $\textrm{Cr/Nb}$(001) film: optimized and uniformly planar strained structures}\label{Sec:S5}
This section deals with another promising prototype candidate, a single layer Cr on Nb(001) substrate, see Fig.~\ref{S5.fig}(a). To the best of our knowledge, no study has been conducted so far on this ultrathin magnetic film sample. The optimized Cr/Mn(001) film exhibits a noncollinear spin texture which is a different type of AFM-SS. Interestingly, the SS ground state is found without any strain. However, we follow the same approach for this system like in Mn/Nb(110) film. In contrast to the Mn/Nb(110) surface u.c., here the surface u.c. possesses square geometry which has $\mathcal{C}_{4v}$ symmetry.

\begin{figure}[H]
	\begin{center}
		\includegraphics[width=1\textwidth]{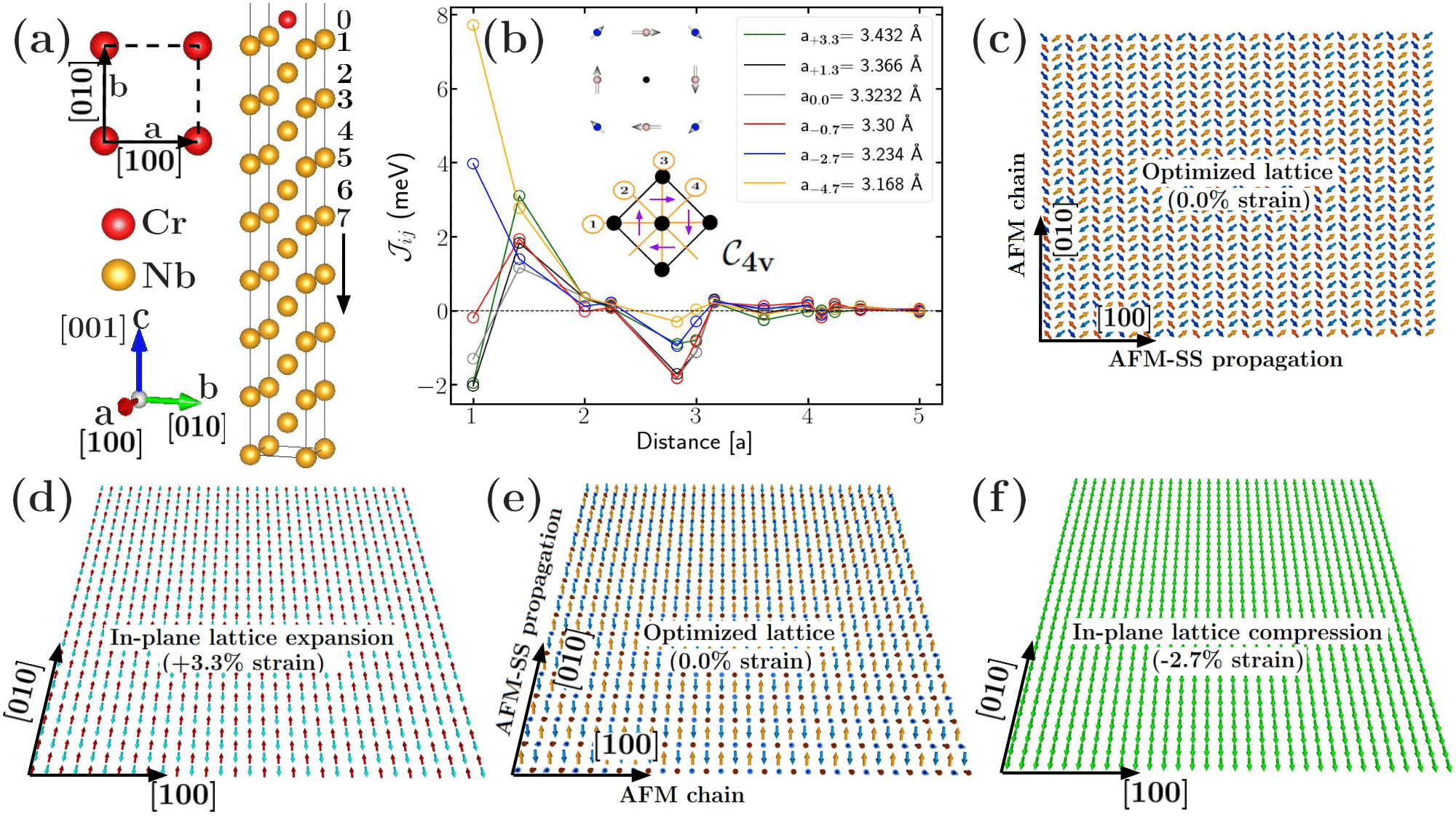}
	\end{center}
	\caption{\hspace*{-0.1cm} In panel (a), we present the unit cell of Cr/Nb(001). The atoms denoted by red and yellow balls are Cr and Nb atoms, respectively. The usual $\mathcal{J}_{ij}'s$ are plotted in panel (b) as a function of distance and the top figure in the inset shows the orientation of NN and NNN DM vectors for the film constructed with optimized lattice constant of Nb. Here, the calculated $\textbf{D}$ vectors are shown on the atoms instead of bonds connection NN and NNN atoms. Below, the sketch of a square surface lattice describing the $\mathcal{C}_{4v}$ symmetry and the mirror planes ($p$=1 to 4) are depicted by yellow lines. The filled black circles represent the surface atoms in the lattice and the microscopic NN $\vect{D}$ vectors (purple arrows) on the crystal surface plane respects the symmetry to point perpendicular to the bond. This is indeed matching with the Moriya rules for $\mathcal{C}_{4v}$ symmetric system~\cite{DMI2}. Panel (c) represents the spin texture with interchanging the [100] and [010] directions. In panels (d), (e) and (f), we demonstrate the evolution of the spin textures driven by the strain parameter. They correspond to the in-plane ferromagnetic (with tensile $+3.3$\% strain), AFM-SS (with zero strain) and in-plane AFM spin texture (with compressive $-3.7$\% strain) respectively.}
	\label{S5.fig}
\end{figure}

In Fig.~\ref{S5.fig}(b), we illustrate the behavior of $\mathcal{J}_{ij}$'s under different strain values and the inset describes the orientation of DMI vectors. Note that, following the $\mathcal{C}_{4v}$ symmetry present in the system, the orientation of DMI vectors around a Cr atom are perpendicular to the bond connecting NN and NNN Cr atoms~\cite{DMI2}. As we keep increasing the strain starting from 3.3 \% tensile, we observe that the NN exchange interaction changes sign \ie a weak AFM coupling becomes a strong FM coupling, see the exchange interaction part in Table~\ref{tab2}. On the other hand, the next dominating exchange constant, the NNN one remains FM for all values of strain. The $\mathcal{D}$ is small here too, owing to the weak SOC strength in Nb. However, compared to Mn/Nb(110) film, the exchange constants are found to be relatively weak. Interestingly, in contrast to Mn/Nb(110), the MCA constants are all in-plane, see the third column in Table~\ref{tab2}. A strong variation 
in $\mathcal{J}$'s under strain makes Cr/Nb(001) film an exciting playground for tailoring magnetism.

Indeed, in Figs.~\ref{S5.fig}(d)-(f), with varying strain, we show that the in-plane AFM ground state for the expanded surface u.c. changes to an in-plane FM ground state under compression. Surprisingly, a noncollinear spin texture (AFM-SS) appears as the ground state with the optimized lattice geometry ($a_\textrm{opt}$= 3.3232 \AA). This particular AFM-SS in Fig.~\ref{S5.fig}(e) exhibits an AFM chain along [100] direction and spins are rotating in the $yz$-plane. So, the AFM-SS is propagating along [010] direction with the sense of spin rotation left-handed (left-handed cycloid). Here, the period of the spiral is small compared to that of Mn/Nb(110), $\lambda_\textrm{Cr}\approx$ 1.33 $\rm{nm}$. This is due to the fast rotation of spins along [010] direction. Here, the interplay between relatively weak, both $\mathcal{J}$'s and $\mathcal{D}$'s stabilizes the SS even in the presence of small in-plane anisotropy. Respecting the $\mathcal{C}_{4v}$ symmetry, the chiral AFM-SS ground state is found to be degenerate in our simulation. The texture in Fig.~\ref{S5.fig}(c) represents another AFM-SS state where the AFM chain along the [010] direction rotates in the $xz$-plane and hence, the propagation direction is [100]. 

\begin{table}[H]
	\centering
	\begin{tabular}{|c|c|c|c|c|c|c|c|c|c|c|c|c|}
		\hline
		Lattice & Magnetic mom. & MCA & \multicolumn{4}{|c|}{DMI magnitude} & \multicolumn{5}{|c|}{Isotropic exchange} & Ground state \\
		constant in \AA & $\mathcal{M}_\textrm{Cr}$ in $\mu_\textrm{B}$ & $\mathcal{K}$ in & \multicolumn{4}{|c|}{in $\rm {meV}$} & \multicolumn{5}{|c|}{interaction ($\rm {meV}$)} & spin texture \\ \cline{4-12}
		(a$_\textrm{strain}$) & KKR-GF,~(VASP) & $\rm{meV}$/Cr & $\mathcal{D}_{01}$ & $\mathcal{D}_{02}$ & $\mathcal{D}_{03}$ & $\mathcal{D}_{04}$ & $\mathcal{J}_1$ & $\mathcal{J}_2$ & $\mathcal{J}_3$ & $\mathcal{J}_4$ & $\mathcal{J}_5$ & \\ \hline
		3.432 (a$_{+3.3}$) & 2.78, (2.83) & -0.40 & 0.32 & 0.02 & 0.02 & 0.1 & -1.96 & 3.1 & 0.32 & 0.07 & -0.91 & In-plane AFM \\ \hline
		3.366 (a$_{+1.3}$) & 2.96, (2.84) & -0.41 & 0.60 & 0.29 & 0.03 & 0.09 & -2.04 & 1.83 & 0.34 & 0.12 & -1.72 & In-plane AFM \\ \hline
		3.3232 (a$_{0.0}$) & 2.69, (2.81) & -0.37 & 0.85 & 0.51 & 0.001 & 0.05 & -1.31 & 1.16 & 0.36 & 0.15 & -1.82 & AFM-ss \\ \hline
		3.300 (a$_{-0.7}$) & 2.69, (2.62) & -0.37 & 0.89 & 0.58 & 0.02 & 0.05 & -0.19 & 1.93 & -0.03 & 0.08 & -1.84 & AFM-ss \\ \hline
		3.234 (a$_{-2.7}$) & 2.79, (2.55) & -0.28 & 0.97 & 0.46 & 0.15 & 0.14 & 3.97 & 1.39 & 0.11 & 0.22 & -0.96 & In-plane FM \\ \hline
		3.168 (a$_{-4.7}$) & 2.38, (2.36) & -0.28 & 0.7 & 0.31 & 0.11 & 0.15 & 7.72 & 2.76 & 0.33 & 0.17 & -0.31 & In-plane FM \\ \hline
	\end{tabular}
	\caption{The first three columns provide the surface lattice constant under strain, $\mathcal{M}_\textrm{Cr}$ obtained from both KKR-GF and VASP in the parenthesis and single ion MCA constant (all are in-plane), respectively. Next, the $\mathcal{D}_{0j}$ and $\mathcal{J}_i$ for first few neighboring cells are tabulated. The different lattice constants signify the system is constructed with different uniform biaxial strain values: from top to bottom, tensile 3.3\% (a$_{+3.3}$), tensile 1.3\% (a$_{+1.3}$), optimized 0.0\% (a$_{0.0}$), compressive 1.6\% (a$_{-0.7}$), compressive 2.7\% (a$_{-2.7}$), and compressive 4.7\% (a$_{-4.7}$). The strains are calculated with respect to the optimized lattice constant, $a_\textrm{opt}$=3.3232 \AA. The last column clearly indicates a transition from an in-plane AFM state to an in-plane FM state via a noncollinear AFM-SS state.}
	\label{tab2}
\end{table}

\begin{figure}[H]
	\begin{center}
		\includegraphics[width=0.8\textwidth]{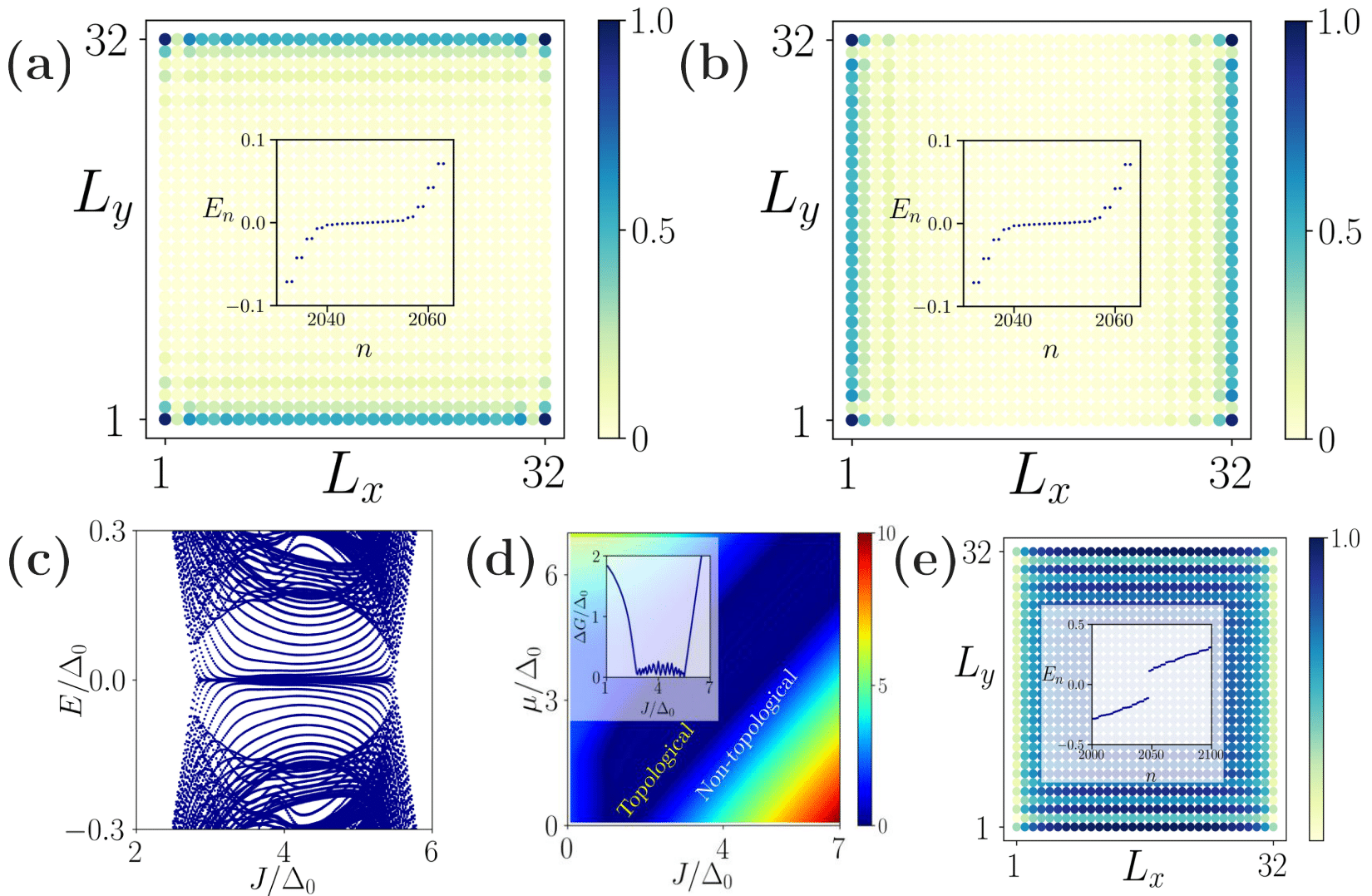}
	\end{center}
	\caption{\hspace*{-0.1cm} By using a square domain within $L_x \mhyphen L_y$ plane, in panels (a) and (b) we depict the normalized LDOS for $E=0$ states at $J= 3.5\Delta_0$, indicating the MFEM at parallel edges along (100) direction (Fig.~\ref{S5.fig}(c), the corresponding spin texture) and parallel edges along (010) direction (Fig.~\ref{S5.fig}(e), the corresponding spin texture), respectively. Insets show the eigenvalue spectrum $E_n$ as a function of the state index $n$. (c) The energy eigenvalues $E$ of the Hamiltonian $H$ (Eq.~(4) in the main text) is plotted as a function of the exchange coupling constant $J$ employing open boundary condition. We choose $\mu=4.0\Delta_0$ and $\Delta_0=t$. (d) The bulk-gap $\Delta G$ profile in the $J \mhyphen \mu$ plane is shown by employing the periodic boundary condition. The dark (light) blue regime represent the TSC phase (non-TSC phase). In the inset plot, we demonstrate the $\Delta G$ as function of $J$ for a fixed $\mu (=4.0\Delta_0)$ for more clarity. (e) The trivial phase occurs for $J= 5.5\Delta_0$ and the eigenvalue spectrum $E_n$ as a function of the state index $n$ clearly shows a gap at $E$=0.} 
	\label{Fig12}
\end{figure}

Now, based on the solutions obtained from the spin lattice model (Eq.~(5) in the main text), Figs.~\ref{Fig12}(a) and (b) describe the LDOS behavior for $E$=0 in the $L_x \mhyphen L_y$ plane for domains with spin textures shown in Fig.~\ref{S5.fig}(c) and Fig.~\ref{S5.fig}(e), respectively. Note, here we consider the spin texture in a domain that is the key ingredient in stabilizing the TSC phase. One can clearly find the edges where localization of the $E=0$ state occurs and in both cases the TSC phase is identified. Additionally, the parallel edges where we find the localized non-dispersive MFEM, strongly depend on the nature of the SS. Particularly, in the present case, the degenerate AFM-SS ground states both stabilize the MFEM on the two parallel edges and depending on the propagation direction, edges are orthogonal to each other, compare Fig.~\ref{S5.fig}(c) (corresponding LDOS in Fig.~\ref{Fig12}(a)) and Fig.~\ref{S5.fig}(e) (corresponding LDOS in Fig.~\ref{Fig12}(b)).  Furthermore, the signature of non-dispersive MFEM is more convincing from the low-lying states (at $E_n=0$) in the eigenvalue spectrum, $E_n$ as a function of the state index $n$, presented in the insets of Figs.~\ref{Fig12}(a) and (b). It is worth to mention that in real experiments, such magnetic domain in general possesses both degenerate solutions~\cite{FerrianiPRL} \ie [100] and [010] propagating SSs together in a single AFM-SS domain and hence, one should expect the MFEM on any edge of a square domain. The calculations are done with exchange coupling strength $J=3.5\Delta_0$. On the other hand, in case of $J=5.5\Delta_0$ (see Fig.~\ref{Fig12}(e)), the system becomes topologically trivial. One can identify the disappearance of the MFEM from the LDOS (at $E=0$) distribution in Fig.~\ref{Fig12}(e) and also, the gapped eigenvalue spectrum in the inset. Thus, the appearance of the MFEM from the real 
material-based study supports our theoretical proposal, see Fig.~3(c) in the main text. 

We depict the energy eigenvalue spectrum ($E/\Delta_0$) as a function of $J/\Delta_0$ employing open boundary condition in Fig.~\ref{Fig12}(c). In comparison to that of Mn/Nb(110) (see Fig.~5(c)), the gapless TSC phase hosting MFEM appears in a wider range of $J$, between 3.5$\Delta_0$ to 5.3$\Delta_0$. To further identify the parameter regime, particularly in the $\mu\mhyphen J$ plane, in which the MFEM appears, we further investigate the bulk-gap $\Delta G=\lvert E_2-E_1\rvert$ employing periodic boundary condition. Here, $E_1$ and $E_2$ represent the two lowest YSR bands. We depict $\Delta G / \Delta_0$ in the $J / \Delta_0 \mhyphen \mu / \Delta_0$ plane in Fig.~\ref{Fig12}(d), see also Fig.~5(d) in the main text for Mn/Nb(110). Here, the gapless TSC regime is also highlighted by the dark blue strip where $\Delta G \simeq 0$ and the regime outside ($\Delta G > 0$) represents the trivial superconducting phase. In the inset of Fig.~\ref{Fig12}(d), we illustrate the bulk-gap $\Delta G$ as function of $J$ for a fixed value of the chemical potential $\mu$, for further transparent visibility of the TSC regime. The bulk-gap $\Delta G$ vanishes in the topological regime ($\sim$ 3.5$\Delta_0$ to 5.3$\Delta_0$) and MFEM appears at the boundary (edges of the sample).

In compared to the Mn/Nb(110) example in the main text and here, for Cr/Nb(001) example we find that the location of the MFEM depends on the propagation direction of the SS. So, owing to the $\mathcal{C}_{4v}$ symmetry in the surface unit cell of Cr/Nb(001), the MFEM can be probed experimentally at any edge of a square domain which generally contains both propagating SSs.

The localized nondispersive MFEMs, extending along a 1D spatial dimension, observed on the two parallel edges as in Figs.~\ref{Fig12}(a) and (b) appear to be highly tunable with the magnetic texture. Our model's unique TSC features open up the possibility of exploring alternating braiding operations in 2D, offering an alternative approach to the conventional zero-dimensional quasiparticles such as Majorana zero modes (MZMs) typically found in 1D systems~\cite{SDSarma2008,BraidingMZM-1}. The presence of controllable underlying magnetic textures enables a highly tunable topological phase, potentially leading to the confinement and transport of MFEM in 2D systems. An intriguing possibility is to engineer a 2D system to create an arrangement of MFEMs selectively localized on the edges of finite domains. Indeed, this concept shares similarities with the idea of braiding Majorana bound states in a magnetic tunnel junction array~\cite{BraidingPRL2016}. Although our model provides a promising platform for manipulating and realizing MFEM in real TM/SC examples, the non-Abelian braiding statistics remain a challenging aspect for next-level topological quantum operations in 2D. Despite the challenges in implementing non-Abelian braiding statistics for MFEM, it can be possible to engineer the edge modes in a manner that allows the injection of a pair of edge vortices into the system. Braiding of an edge vortex might be achievable when it crosses the branch cut of a bulk vortex. In a similar context, a Josephson junction geometry has been explored to inject isolated vortices into chiral edge channels in 2D~\cite{BraidingPRL2019}. However, it is essential to carefully investigate whether the injection of vortices is feasible through flat Majorana edge modes, particularly in the case of vanishing group velocity. Another device proposal in 2D involves a junction of Quantum Anomalous Hall Insulator (QAHI) and topological superconductor (TS) layers, where a properly designed QAHI$-$TS$-$QAHI junction~\cite{BraidingPNAS2018} could serve as an implementation of topologically protected quantum gates. Our model, featuring nondispersive MFEMs, distinguishes itself from existing 2D examples that primarily rely on chiral dispersive edge modes in various proposals. Nevertheless, to fully understand the potential for braiding operations and non-Abelian statistics with nondispersive MFEMs, further investigation and new developments are required.

In summary, these studies with a wide variation in noncollinear AFM spin textures firmly establish the non-trivial TSC phase hosting MFEM, allowing a formal connection between noncollinear magnetism in real materials and model studies. 
\section{YSR/Shiba energy band formation due to the AFM-SS state in the Mn layer on top of $\textrm{Nb}$(110) surface}\label{Sec:S6}

\begin{figure}[H]
	\begin{center}
		\includegraphics[width=0.8\textwidth]{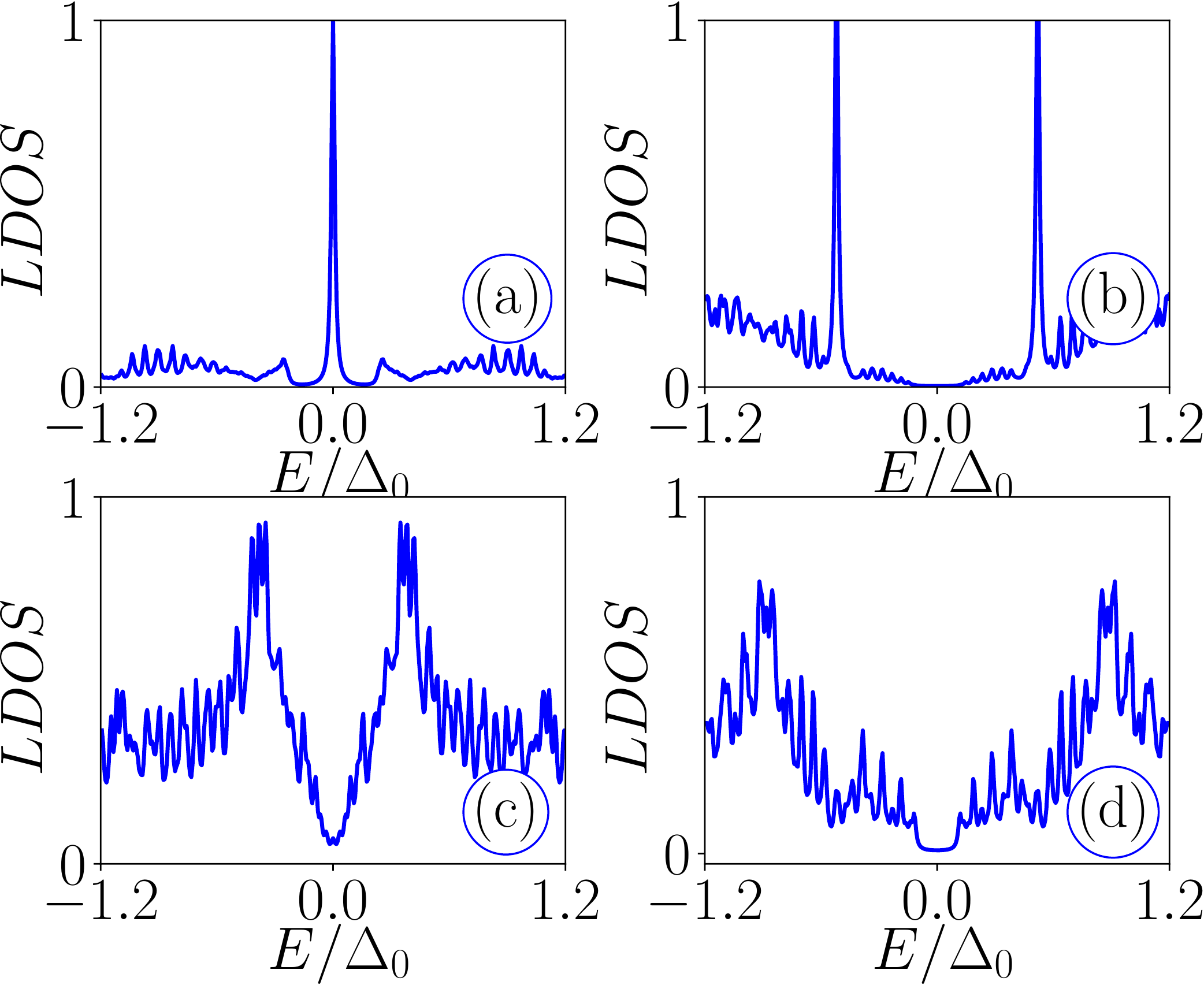}
	\end{center}
	\caption{\hspace*{-0.1cm} Panels (a), (b) correspond to the behavior of $\rm LDOS$ as a function of $E/\Delta_0$ for two different values of exchange coupling strength $J=4.5\Delta_0$ and $J=5.0\Delta_0$ respectively, when we measure $\rm LDOS$ at the edge of the system. Panels (c), (d) indicate the formation of Shiba bands at the same parameter values when we measure $\rm LDOS$ 
		at the middle of the system.} 
	\label{Fig13}
\end{figure}

In this section of the SM, we discuss about the non-topological Shiba band formed within the energy scale $E= \pm\Delta_0$ in the presence of Mn layer as magnetic impurities (AFM-SS) placed on top of an $s$-wave superconductor Nb (110). We obtain the signature of Shiba band when we compute $\rm LDOS$ at the middle of the system. These Shiba bands play the pivotal role during the topological phase transition.

In Fig.~\ref{Fig13}(a), we obtain a sharp peak at $E=0$ in the LDOS behavior for $J=4.5\Delta_0$. This is a signature of the MFEM (topologically non-trivial) when calculated at the edges of the system. On the other hand, in Fig.~\ref{Fig13}(c) we depict the corresponding YSR (or simply Shiba) band features in LDOS with the same exchange coupling value $J=4.5\Delta_0$, when we compute $\rm LDOS$ at the middle of the system. The semimetallic behavior arising in the LDOS [see Fig.~\ref{Fig13} (c)] is the signature of the gaplessness (graphene-like behavior) of the bulk YSR band in the topological regime. This is consistent with the Fig.~2(d) (blue curve) in the main text. On the other hand, Fig.~\ref{Fig13}(b) and Fig.~\ref{Fig13}(d) represent $\rm LDOS$ at the edge and middle (YSR/Shiba band) of the system respectively in the non-topological regime (when $J=5.0\Delta_0$), where MFEM peak disappears at the edges, see Fig.~\ref{Fig13}(b). Concomitantly, LDOS in Fig.~\ref{Fig13} (d) exhibits a gapped superconductor instead of a semimetallic phase. Similarly, in Fig.~\ref{Fig14}, panels (a), (b), (c), (d) correspond to the spatial variation of the $\rm {LDOS}$ at finite energy $E=0.1\Delta_0$ in the $L_x \mhyphen L_y$ plane for four different values of the exchange coupling strength $J= 4.0\Delta_0, 4.2\Delta_0, 4.5\Delta_0, 5.0\Delta_0$ respectively. One can observe a significant variation of the non-topological Shiba bands for each case due to the presence of the Mn layer as magnetic SS.


\begin{figure}[H]
	\begin{center}
		\includegraphics[width=0.45\textwidth]{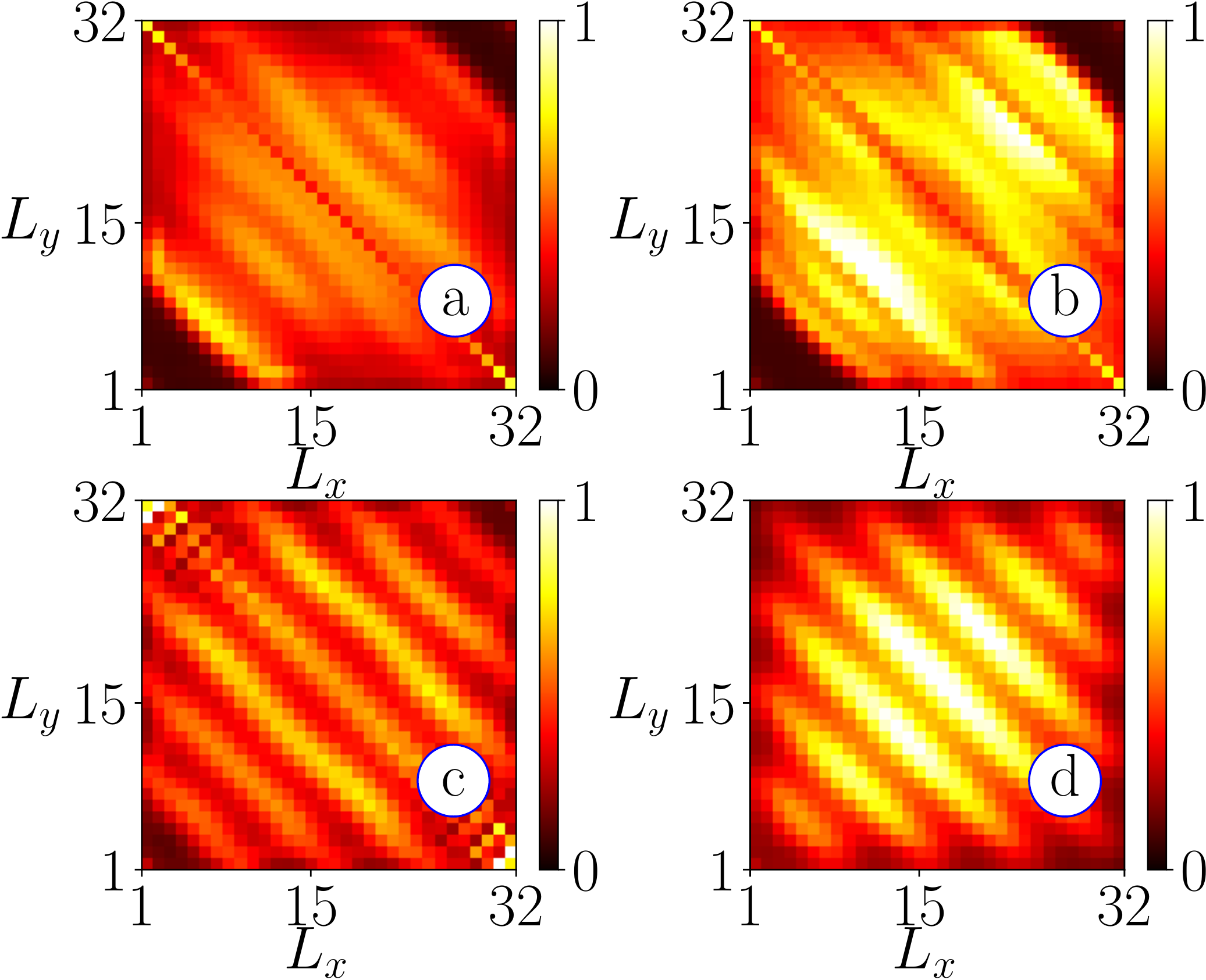}
	\end{center}
	\caption{\hspace*{-0.1cm} Panels (a), (b), (c), (d) correspond to the vatiation of $\rm LDOS$ at finite energy $E=0.1\Delta_0$ in $L_x \mhyphen L_y$ plane (highlighting corresponding features of 
		YSR band) with respect to the exchange field strength $J= 4.0\Delta_0, 4.2\Delta_0, 4.5\Delta_0, 5.0\Delta_0$ respectively.}  
	\label{Fig14}
\end{figure}

\section{Derivation of an effective $p$-wave superconducting pairing}

In the following section, we investigate the characteristics of the effective pairing mechanism~\cite{Masatoshi2009,Masatoshi2010}  that leads to the formation of the gapless topological superconducting phase in 2D, hosting nondispersive MFEM in our model. Our analyses begin with the consideration of the effective $\mathbf{k}$-space model, 
\begin{eqnarray}
	\tilde{H} (\vect{k})\!=\!&\xi_{\vect{k},\vect{g}}\tau_{z}\! +\! \frac{1}{2} \vect{g} \! \cdot \! \vect{k} \ \sigma_z\tau_{z} \! +\! J\sigma_x \! + \! \Delta_{0}\tau_{x} \ .  
	\label{eq1}
\end{eqnarray}
This is also presented in Eq.~(3) in the main text of our study, where all terms are appropriately defined.
The Bogoliubov de-Gennes (BdG) form of the above Hamiltonian reads,
\begin{eqnarray}
	\tilde{H} (\vect{k})=\begin{pmatrix} 
		\xi_{\vect{k},\vect{g}}\sigma_{0}+J\sigma_x+\! \frac{1}{2} \vect{g} \! \cdot \! \vect{k}\sigma_z & \Delta_{0}\sigma_{0} \\
		\Delta_{0}\sigma_{0} & -\xi_{\vect{k},\vect{g}}\sigma_{0}+J\sigma_x-\! \frac{1}{2} \vect{g} \! \cdot \! \vect{k}\sigma_z\\
	\end{pmatrix}, 
	\label{eq2}
\end{eqnarray}
where, $\xi_{\vect{k},\vect{g}}$=$\frac{1}{2}\left(k_x^2+k_y^2+\!\vect{g}^2 \right)\!-\!\mu$,  $\vect{k}=(k_x,k_y)$ and $\vect{g}=(g_x,g_y)$. We assume that $g_x=g_y=g$. 
In our quest to establish a dual representation for the BdG Hamiltonian, as described in Eq.~(\ref{eq2}), we aim to find an equivalent dual Hamiltonian denoted as $\tilde{H}^D(\vect{k})$. In this process, we utilize a unitary operator responsible for the duality transformation. We define the selected unitary operator for this purpose as follows:
\begin{eqnarray}
	S=\frac{1}{\sqrt{2}}\begin{pmatrix} 
		1_{2\times 2} & 1_{2\times 2} \\
		1_{2\times 2} & -1_{2\times 2}\\
	\end{pmatrix}. 
	\label{eq3}
\end{eqnarray}
By this unitary transformation, the Hamiltonian becomes,
\begin{eqnarray}
	\tilde{H}^D (\vect{k})= S^{\dagger}\tilde{H} (\vect{k})S\ .
	\label{eq4}
\end{eqnarray}
Hence, the equivalent dual Hamiltonian can be written as,
\begin{eqnarray}
	\tilde{H}^D (\vect{k})=\begin{pmatrix} 
		\epsilon_{\vect{k}}^{D}\sigma_0+J\sigma_x &  \tilde{\Delta}^D (\vect{k})\sigma_0\\
		\tilde{\Delta}^D (\vect{k})\sigma_0 & -\epsilon_{\vect{k}}^{D}\sigma_0+J\sigma_x\\
	\end{pmatrix}, 
	\label{eq5}
\end{eqnarray}
where,
\begin{eqnarray}
	&&\epsilon_{\vect{k}}^{D}=\Delta_0\ , \non\\
	&&\tilde{\Delta}^D (\vect{k})=\frac{1}{2}g(k_x+k_y)\sigma_z+\xi_{\vect{k},\vect{g}}\sigma_{0} \ .
	\label{eq6} 
\end{eqnarray}
At low energy limit ($k_x,k_y\sim0$), keeping the linear order of $k_x$ and $k_y$, the effective pairing takes the form,
\begin{eqnarray}
	\tilde{\Delta}^D (\vect{k})\simeq \left(\mu-\frac{g^2}{2}\right)\sigma_{0}+\frac{g}{2}(k_x+k_y)\sigma_z\ .
	\label{eq7}
\end{eqnarray}
In case of spin up ($\uparrow\uparrow$)  and spin down ($\downarrow$$\downarrow$) channel dual pairing becomes,
\begin{eqnarray}
	\tilde{\Delta}_{\uparrow/\downarrow}^D (\vect{k})= \left(\mu-\frac{g^2}{2}\right)\pm\frac{g}{2}(k_x+k_y)\ .
	\label{eq8}
\end{eqnarray}
Thus, it is evident that the spin-orbit coupling term ($\frac{1}{2}\vect{g}.\vect{k}\sigma_z$), arising from the spin spiral in the original Hamiltonian $\tilde{H} (\vect{k})$, is formally transformed into an effective ``$p_x + p_y$-type" pairing with a constant kinetic energy term $\epsilon_{\vect{k}}^{D}=\Delta_0$ in the dual Hamiltonian. This intriguing result suggests that the signature of the topological superconducting phase, namely the nondispersive MFEM, arises due to a unique $p$-wave pairing mechanism resulting from the interplay between the spin-spiral state and the normal $s$-wave superconductivity. The unitary transformation $S$, which is independent of $\vect{k}$, does not alter the topological invariant, ensuring the topological equivalence between Eq.~(\ref{eq2}) and Eq.~(\ref{eq5}).

\section{Effect of Rashba SOC on single-orbital model and an extension to multi-orbital theory}

In the presence of single-orbital (SO) channel, Rashba SOC term becomes~\cite{Ojanen12015},
\begin{eqnarray}
	H_{{\rm{SO}}}^R=\lambda_R\tau_z(\sigma_x\sin k_y-\sigma_y\sin k_x) \ ,
	\label{eq9}\ 
\end{eqnarray}
where $\lambda_R$ is the Rashba SOC strength. Here, the Pauli matrices {$\sigma$} and {$\tau$} operate on the spin and particle-hole degrees of freedom, respectively. Similarly, in the presence of multi-orbital (MO) channels the same term [Eq.~(\ref{eq9})] added in the Hamiltonian [Eq.~(4) in the main text] takes the form~\cite{HughesTaylor2008,AndersMathias2016},
\begin{eqnarray}
	H_{{\rm{MO}}}^R=\lambda_R\tau_z\Gamma_{\textrm{orb}}(\sigma_x\sin k_y-\sigma_y\sin k_x)\ ,
	\label{eq10}\ 
\end{eqnarray}
where, $\Gamma_{\textrm{orb}}=\begin{pmatrix} 
1 & 0 \\
0 & 0 \\
\end{pmatrix}$ represents orbital degrees of freedom. Therefore, we study the effect of MO and SOC on prototype heterostructures where the magnetic layer is based on 3d-transition metal (TM) element deposited on the $s$-wave superconducting (SC). 

\begin{figure}[H]
	\begin{center}
		\includegraphics[width=1.02\textwidth]{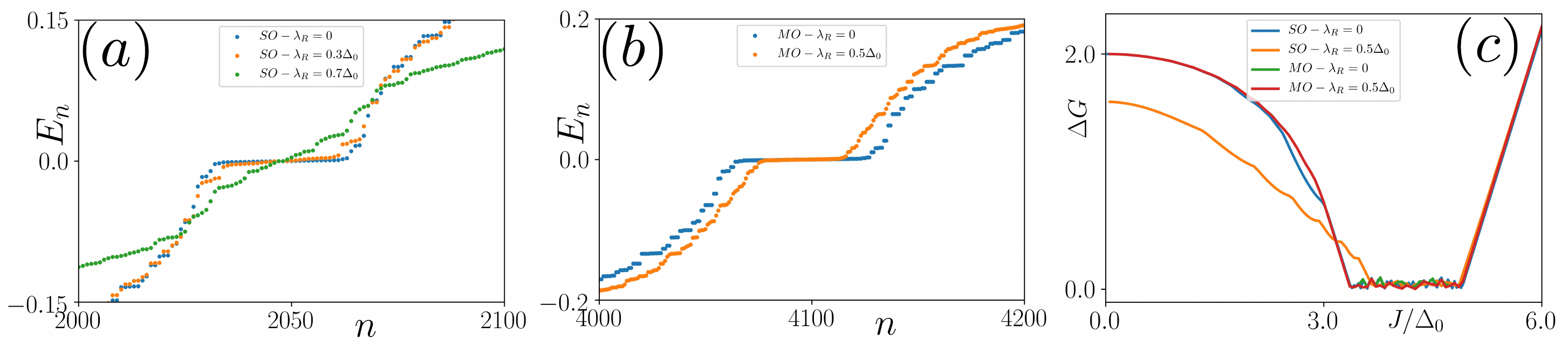}
	\end{center}
	\caption{\hspace*{-0.1cm} Panels (a) and (b) depict the eigenvalue spectra of the lattice Hamiltonian containing SO and MO channels, respectively. Here, we have varied the Rashba SOC strength, $\lambda_R$, while using an open boundary condition in the 2D finite domain. The lattice model takes into account the noncollinear spin texture in the form of AFM-SS (strained Mn/Nb(110) system). The exchange coupling strength is chosen as $J=4.5\Delta_0$ for all cases. Panel (c) illustrates the bulk gap as a function of the exchange coupling strength, while considering periodic boundary conditions in 2D. We investigate four different scenarios: (i) SO without Rashba SOC, (ii) SO with Rashba SOC, (iii) MO without Rashba SOC, and (iv) MO with Rashba SOC. The additional model parameter values are chosen as $\mu=4.0\Delta_0$ and $t=\Delta_0$. }
	\label{fig1}
\end{figure}

In Fig.~\ref{fig1}(a) and (b), we depict the eigenvalue spectrum in the topological phase for SO and MO cases [at the level of tight-binding model Eq.~(4) of our main text], respectively. The strength of the Rashba SOC in both cases has been varied to illustrate its impact on our results. The qualitative similarity of the overall physics in the presence of moderate strength Rashba SOC is evident from the results presented in Fig.~\ref{fig1}(a) for the SO model. However, for strong Rashba SOC strength ($\lambda_{R}=0.7\Delta$), the MFEM becomes relatively dispersive without altering the topological nature of the TSC phase. Incorporating additional orbital channels in the MO model and including the SOC term (Eq.~\ref{eq10}) does not alter the existence of MFEMs in the topological phase. We observe only slight overall renormalization in the gapless topological superconducting phase hosting nondispersive MFEMs. Indeed, the fundamental physics of MFEMs remains qualitatively similar even with the inclusion of various basic ingredients such as multi-orbital channels and the Rashba SOC term.
In Fig.~\ref{fig1}(c), we demonstrate the bulk gap ($\Delta G$) profile as a function of exchange coupling strength $J$ for four different scenarios: (i)~SO without Rashba SOC, (ii)~SO with Rashba SOC, (iii)~MO without Rashba SOC, (iv)~MO with Rashba SOC.Notably, the topological gapless phase is observed to exist approximately between $J=3.5\Delta_0$ and $J=4.7\Delta_0$ for all the scenarios mentioned above. Hence, the presence of additional orbital channels and the effect of SOC has minimal influence on the overall physics of MFEMs. 
\vspace{-0.7cm}
\begin{figure}[H]
	\begin{center}
		\includegraphics[width=1.0\textwidth]{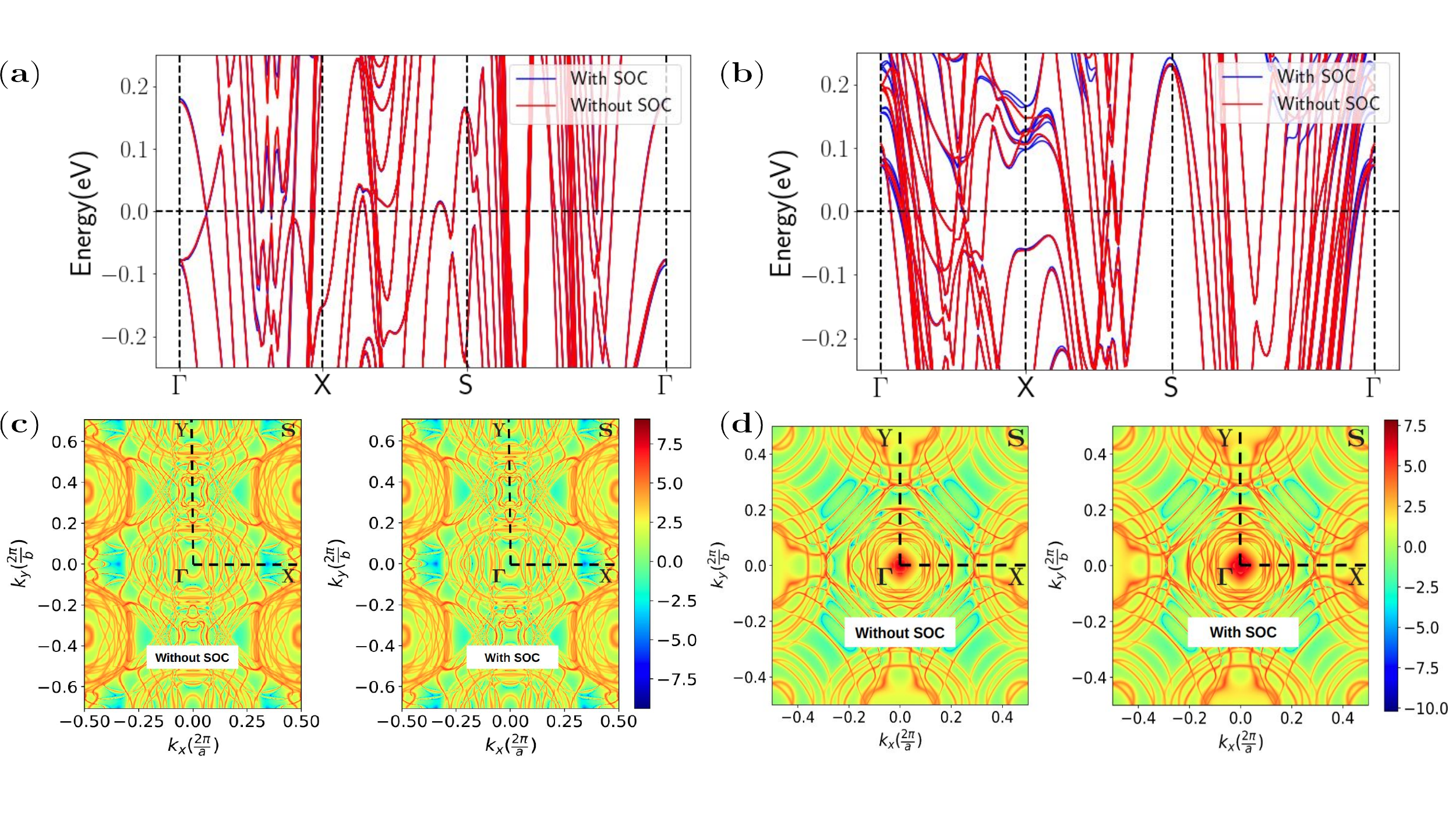}
	\end{center}
	\caption{\hspace*{-0.1cm} The panels labelled as (a) and (b) display the band structures obtained from the \textit{ab~initio} electronic structure calculations for the Nb(110) and Nb(001) surfaces, respectively. These band structures are computed with and without taking into account the effect of SOC. Furthermore, the panels labelled as (c) and (d) display the energy contours of the Fermi surface for both surfaces. Similar to the band structure panels, these Fermi surface energy contours are also computed with and without considering SOC.}
	\label{fig2}
\end{figure}

In the main text, we opted for the minimal model without including the SOC term to provide a clear and unambiguous exploration of the real example systems. The choice was based on the empirical observation that the real example systems we studied are not significantly influenced by the Rashba SOC term, which may observe in surface/interface systems. In Fig.~\ref{fig2}, we have presented the band structure and the Fermi surface plots for $s$-wave superconductor, Nb(110) and Nb(001) in slab geometries. The influence of the SOC effect on the band structure is minimal, as evidenced by the negligible shifts in the blue and red bands in Fig.~\ref{fig2}(a) and (b) for Nb(110) and Nb(001), respectively. Moreover, the Fermi surface plots for both cases, with and without SOC, shown in Fig.~\ref{fig2}(c) and (d) for Nb surfaces, clearly demonstrate the minimal impact of the SOC effect.
\end{onecolumngrid}
\end{document}